\documentstyle[12pt,aasms4]{article}
\begin{document}

\title{Photometric Periodicities of Be/X-ray Pulsars \\ in the Small
Magellanic Cloud\footnote{This paper utilizes public domain data obtained
by the MACHO Project, jointly funded by the US Department of Energy
through the University of California, Lawrence Livermore National
Laboratory under contract No.\ W-7405-Eng-48, by the National Science
Foundation through the Center for Particle Astrophysics of the University
of California under coopertative agreement AST-8809616, and by the Mount
Stromlo and Siding Spring Observatory, part of the Australian National
University. } } 

\author{ P.C.\ Schmidtke \& A.P.\ Cowley }

\affil{Dept.\ of Physics \& Astronomy, Arizona State Univ., Tempe, AZ,
85287-1504, USA; paul.schmidtke@asu.edu, anne.cowley@asu.edu } 

\begin{abstract}

Analysis of the longterm photometric variability of 7 Be/X-ray-pulsar
systems in the Small Magellanic Cloud has been carried out.  We find
a variety of types of variability are present, including: longterm
irregular changes, periodic orbital outbursts due to interaction between
the stars (weeks to months), low-amplitude quasi-periodic variations of
the equatorial disk (days), and nonradial pulsations of the Be primary
star (hours).  This is the first time nonradial pulsations have been
identified in Be/X-ray binaries, although they were previously known in
some single Be stars. 
 
\end{abstract}

                                    
\section{Introduction}

Very luminous, high-mass X-ray binaries (HMXB) are generally divided into
two groups: 1) those containing a Roche-lobe filling supergiant which
transfers mass onto a neutron star or black-hole companion, and 2) wider
binaries containing a Be star and a neutron star which accretes material
from the equatorial disk of the primary.  In the latter systems the
compact star is often found to be an X-ray pulsar.  In the Galaxy the HMXB
are about equally divided between these two broad classes, but in the
Magellanic Clouds the Be/pulsar systems make up the majority of HMXB.  In
the SMC we recognize only one supergiant system (SMC X-1), while several
dozen X-ray pulsars with Be-star companions are known (e.g. Haberl \&
Sasaki 2000; Coe et al.\ 2002). 

Only a few of the Be/neutron star systems have been studied in detail, but
from these it appears the stars are in fairly wide but eccentric orbits.
Hence the two stars interact primarily near periastron passage when the
neutron star may enter the extended equatorial disk of the Be star and
experience increased accretion of gas.  This gives rise to both X-ray and
optical outbursts from which the orbital period can be determined.  Corbet
(1984) and others (e.g. in't Zand et al.\ 2001; Charles \& Coe 2004) have
shown that the pulse and orbital periods in such Be/X-ray systems are
related.  This correlation was explained by Waters \& van Kerkwijk (1989)
as resulting from the slow equatorial Be winds which affect the equilibrium
spin period of the pulsar. 

A notable example of periodic orbital outbursts is seen in the LMC source
A0535$-$668, where the eccentricity is $\sim$0.7.  This system shows
extreme changes in spectrum and brightness at all wavelengths near
periastron (e.g. Charles et al.\ 1983; Hutchings et al.\ 1985).  For many
of the other Be/X-ray systems we see less dramatic but clear periodic
outbursts (e.g. AX~J0049.4$-$7323, Cowley \& Schmidtke 2003), although the
orbital eccentricity has not yet been determined for almost all of these
systems.  Schmidtke et al.\ (2004) identified new orbital periods in 7
Magellanic Cloud Be/X-ray systems based on longterm MACHO and OGLE-II
photometric data.  In addition to orbital outbursts, they found
quasi-periodic variations (QPV) which they suggested were due to
reverberations in the Be star's equatorial disk following penetration by
the neutron star.  Generally the QPV have timescales of days.  The QPV
periods generally differ by a few percent following each orbital
interaction. 
  
As pointed out by Schmidtke et al.\ (2004, and references therein)
Be/neutron star binaries often show longterm light curves with
considerable variability.  Some exhibit prominent `swoops' or irregular
outbursts (several examples are shown in Fig.\ 4 of Coe et al.\ 2005),
while others show a lot of scatter in the light curve.  For a few, their
periodic behavior is clearly visible without any analysis (e.g.
RX~J0049.4$-$7323, Cowley \& Schmidtke 2003).  However, for most systems a
full analysis of the photometry is needed to identify any periodic signal.

In the present paper we have investigated 7 more SMC systems in an effort
to better understand the variability seen in Be/X-ray-pulsar systems.  The
particular systems were selected on the basis of several criteria. First,
they have well determined positions, and hence we are confident in the
optical identifications.  Second, the sources had to be included in either
the MACHO or OGLE surveys.  Third, we chose sources where the variability
exceeds the photometric errors.  Additionally, the OGLE sources are all in
their ``Difference Image Analysis" (DIA) catalogue", so they had already
been selected as variables. 

In addition to orbital outbursts and QPV, we have discovered that some of
these Be stars also show nonradial pulsations (NRP) with timescales of
hours to over a day.  In retrospect, this is not surprising, since it is
known that some single Be stars show such variability (e.g. Balona,
Sterken, \& Manfroid 1991; Balona 1992; Balona \& James 2002; Percy,
Harlow, \& Wu 2004). 

\section{Analysis of Optical Data from the MACHO and OGLE-II Projects} 

Our photometric data come from two large surveys which allow public access
to their data: the OGLE-II survey (Udalski, Kubiak, \& Szymanski 1997;
Zebrun et al.\ 2001) and the MACHO survey.  OGLE-II provides photometry in
$I$ for variables in their DIA catalogue.  The MACHO site gives `blue' and
`red' instrumental magnitudes for all stars in their fields, and these
data can be transformed to standard $V$ and $R$ colors (Alcock et al.\
1999).  (Note: For users looking at the web display of `calibrated' MACHO
light curves for SMC sources, one must add 0.75 mag to the scale shown,
since only half of the actual integration time (300 s versus 600 s) is
used when generating these quick-look light curves.) 

Table 1 gives the equinox 2000 MACHO position of the optical stars, the
mean $R$ magnitudes, the mean $V-R$ colors, the MACHO and OGLE-II
identification numbers, and the catalogue numbers from ``H$\alpha$-bright
Stars in the SMC" (Meyssonnier \& Azzopardi 1993; hereafter called MA93)
for the systems we studied. 

Each data set ($V$, $R$, $I$) was flattened, or pre-whitened, to remove
longterm trends.  This was done by fitting a low-order polynomial to long
stretches of data in each color, excluding extreme outlying points or
sections of the light curve where unrepeated sudden changes occurred.  The
resulting ``detrended" magnitudes ($V^*$, $R^*$, $I^*$) were then
analyzed.  In some cases we separated the MACHO data by east-pier and
west-pier designations prior to detrending in order to remove any
systematic differences between the two instrumental configurations. 

Using the method described by Horne \& Baliunas (1986), we searched for
periodic behavior in the detrended light curves.  The resulting
periodograms are shown for each system discussed below.  In cases where
the periodic light curves are very non-sinusoidal, this method of analysis
may display only low power at the true period, since it searches for
sinusoidal variations.  Usually aliases of the true period are stronger,
as is the case for several of the systems studied in this paper.  In these
cases the period is better revealed using the phase dispersion
minimization (PDM) technique described by Stellingwerf (1978).  The PDM
method can identify periodicities in light curves having an arbitrary but
repeatable shape, such as the recurring outbursts seen in some of the
systems described here.  More details are given with each source where the
PDM method has been used.  Typically, we searched for periods in the range
0.25-1000 days, covering a minimum of 10$^4$ test frequencies. 

For a given periodogram, the significance of its peaks was explicitly
assessed by creating 1000 data sets, with randomized magnitudes and dates.
From a simple tally of the strongest power peak within each of these sets,
we determined the 99\% confidence level.  That is, in only 1\% of the
random data sets did the peak power exceed the stated 99\% level.  Nearly
all of the periods found in this study are significant at or above this
value.  This assessment assumes that the signal is strictly sinusoidal and
coherent.  If these conditions do not hold, such as in QPV or NRP, then
the observed power can be somewhat less than the ideal case, so that the
significance of the peak {\it appears} to be lower. 

Most of the figures display folded light curves, using the ``detrended"
magnitudes.  Hence, these light curves no longer contain information about
the brightness of the system.  However, the true magnitudes are plotted in
the longterm light curves. 

\section{Individual Be/X-ray Pulsar Systems}

\subsection{ XMMU~J004723.7$-$731226 = RX~J0047.3$-$7313 }

This 263.6 s X-ray pulsar was discovered by Ueno et al.\ (2004) but was
also studied by Haberl \& Pietsch (2004).  Because of its very good {\it
XMM-Newton} position, we can confidently identify it with MACHO
212.15792.77.  It also appears to be MA93\#172, confirming the primary as
an emission-line B star. 

The longterm MACHO $V$ light curve is shown in the upper panel of Fig.\ 1.
Both the MACHO and OGLE-II data show a general scatter of $\sim$0.1 mag
superimposed on a slow brightening trend.  This system is one of the cases
where our standard analysis reveals aliases rather than the fundamental
period.  The periodogram from the $I^*$ data shows power at 12.3 d and 9.8
d, but light curves folded on these periods are very scattered.  The PDM
analysis of the $I^*$ data shows that the fundamental period is
P=49.1$\pm$0.2 days (third panel of Fig.\ 1), which we assume is the
orbital period.  However, both the periodogram and the PDM variance for
the $V^*$ and $R^*$ data show no significant periods.  In similar Be/X-ray
systems the orbital outbursts are usually strongest in $I$, and hence the
lack of periodic variation in $V$ or $R$ is not surprising.  Edge et al.\
(2005a) report P=48.8$\pm$0.6 d, also from OGLE $I$ data, although details
of their analysis are not yet available. 

The folded orbital $I^*$ and $V^*$ light curves are shown in the bottom
panel of Fig.\ 1.  In $I^*$ there is a small outburst lasting for
$\sim$0.1P, with an amplitude of $\sim$0.02 mag.  Examining different
segments of the data reveals that from cycle to cycle the shape of the
light curve is similar, but some outbursts are stronger than others.   
For a short time before and after each outburst the source appears to be
$\sim$0.01 mag fainter than the mean.  There is no corresponding outburst
in $V^*$ or $R^*$. 

\subsection{ RX~J0049.1$-$7250 = AX~J0049$-$729 }

X-ray pulses at 74.7 s from RX~J0049.1$-$7250 were discovered by Corbet et
al.\ (1998) and confirmed by Yokogawa \& Koyama (1998a).  The optical
counterpart is identified with the emission line Star \#1 of Stevens, Coe,
\& Buckley (1999).  The system has a mean magnitude of $R\sim$16.9, and
its longterm light curve shows variability of $\sim$0.2 mag (see Fig.\ 2).

MACHO light curves in $V$ and $R$ were analyzed in two segments (A and B,
as marked on Fig.\ 2).  Both the OGLE-II and MACHO photometric data show a
highly significant periodicity at 33.4$\pm0.4$ days, with its detection
being well over the 99\% confidence level.  This period is likely to be
the orbital period, and it is in good agreement with the relation between
orbital and pulse periods first recognized by Corbet (1984).  The second
panel of Fig.\ 2 shows the periodogram of $R^*$ data taken from segments A
and B.  The $R^*$ light curve folded on P=33.4 days, is shown in the third
panel.  In $V^*$ light, the 33-day period is present but weaker. 

The periodogram also reveals several significant peaks with periods in the
range of 2.39-2.40 days which we identify as quasi-periodic variations of
the Be star's disk.  The QPV have the dominant power in $V^*$ where the
33-day signal is weaker.  The $R^*$ and $V^*$ data from Seg A, folded on
the QPV 2.4-day period, are plotted in the bottom panel of Fig.\ 2. 

The longterm $I$ light curve for RX~J0049.1$-$7250 is shown in the upper
panel of Fig.\ 3.  Note the super-outburst near MJD 50760.  The middle
panel displays the periodogram from $I^*$ data, with the observations near
the super-outburst excluded.  The 33.4-day period is prominent.  The bottom
panel shows the $I$ data folded on this period, with the super-outburst
superimposed in a different symbol (open circles).  The mean amplitude of
the 33-day variation is $\Delta$$I\sim$0.03 mag.  The super-outburst is in
phase with the other periodic brightenings -- it just happened to be much
larger, perhaps due to denser material surrounding the Be star at the time
of the interaction. 

We note that Coe \& Orosz (2000) reported that they found no periods in
the OGLE-II data in the range 1-50 days.  Laycock et al.\ (2004) suggested
a possible orbital period of 642$\pm$59 days based on the intervals
between three strong X-ray outbursts observed with RXTE.  However, with
more X-ray data Galache (2005, private communication) has found
P$_X$$\sim$65.1 days, which may be double the 33.4-day optical period
identified here.  This suggests that a strong X-ray outburst is not seen
at every periastron passage. 

\subsection{ XTE~J0052$-$725 }

The position of this 82.4 s pulsar was determined using $Chandra$ data by
Edge et al.\ (2003).  The longterm $R$ light curve shows the mean
magnitude was fairly constant until MJD 50800, with a scatter of $\sim$0.1
mag.  After that date the source slowly declined by $\sim$0.3 mag in $B$,
$V$, and $I$.  Both $R$ and $I$ longterm light curves are plotted in Fig.\
4.  Coe et al.\ (2005; their Fig.\ 4) using both OGLE-II and -III data
show that the source brightened slightly between MJD 51500 and 52000, but
then fell rapidly by at least 0.6 mag. 

The MACHO $V^*$ and $R^*$ light curves were analyzed in four time segments
(A-D), as shown in Fig.\ 4.  A strong periodic variation, with P=1.328
days, is present in both MACHO colors.  The $R^*$ power spectrum for
segments A through C is shown in the middle panel of Fig.\ 4; the primary
peak and its aliases are marked.  Examining the individual segments, the
power steadily dropped until there was virtually no periodic signal in Seg
D (see bottom panel of Fig.\ 4).  The period remained essentially constant
while the variation was present. 

As expected, the $I^*$ data which overlaps Seg D (from MJD $\sim$50500 to
$\sim$51500) also shows no periodic signal.  However, in spite of the
small amount of $I$ data in Seg E, a power spectrum analysis reveals the
original period returned during this segment, although its phasing
differed by about a quarter of a cycle.  Such a short period with variable
amplitude suggests that it is due to nonradial pulsations of the Be star.
This is the first time NRP have been identified in a Be/X-ray binary,
although similar short periods are observed in some single Be stars (e.g.
Percy et al. 2004; Balona \& James 2002; Stefl \& Balona 1996). 

$R^*$ and $V^*$ light curves from Seg A, B, and C, folded on P=1.328 days,
are shown in Fig.\ 5.  In a given segment, both colors have similar
amplitudes.  The greatest amplitude occurred in Seg A ($\Delta$m$\sim$0.04
mag).  In subsequent segments the amplitude decreased
($\Delta$m$\sim$0.02-0.03 mag in Seg B; $\Delta$m$\sim$0.01 mag in Seg C)
until there was no obvious periodic variation in Seg D.  The $I^*$ light
curve from Seg E, folded on the mean period from Seg A and B, is shown in
the bottom panel of Fig.\ 5.  The amplitude is similar to that in Seg A,
but the phasing differs as noted above. 

We searched for but were unable to find any photometric signature of an 
orbital period up to P=1000 days. 

\subsection{ CXOU~J005455.6$-$724510 = RX~J0054.9$-$7245 = 
AX~J0054.8$-$7244 }

CXOU~J005455.6$-$724510 appears to have been independently discovered as a
$\sim$500 s pulsar by Edge et al.\ (2004a) using $Chandra$ archival data
and by Haberl et al.\ (2004a) from {\it XMM-Newton} data.  Edge et al.\
(2004b) give the pulse period as 503.5$\pm$6.7 s, while Haberl et al.\
(2004b) find P$_{pulse}=499.2\pm$0.7 s. 

The highly accurate $Chandra$ position allows one to confidently identify
it as the emission-line star MA93\#809 and MACHO 207.16254.16.  Although
the field is covered by OGLE-II data, the star is not listed as a variable
in the DIA catalogue.  This is probably due to the fact that the amplitude
of the photometric variation is quite small. 

The longterm $R$ light curve of CXOU~J005455.6$-$724510 is plotted in
Fig.\ 6.  Both a periodogram and the variance from PDM analysis were used
to study the $R^*$ and $V^*$ data.  The $R^*$ data show significant power
at P=273$\pm$6 days (see middle panel in Fig.\ 6; the 273-day peak is well
over the 99\% confidence level).  We infer that this is the orbital
period.  The peak near 0.002 d$^{-1}$ is not seen in the PDM analysis and
therefore is likely to be spurious.  The $R^*$ light curve folded on P=273
days shows a small outburst ($\Delta$$R$$\sim$0.02 mag) with a rapid rise
and slower decline (see bottom panel).  The $V^*$ data show no significant
periods, but as pointed out above for XMMU~J004723.7$-$731226, orbital
outbursts tend to be strongest in $I$, weaker in $R$, and may not be
present in $V$.  Folding the $V^*$ data on 273 days shows that if there is
any outburst in this color, its amplitude is $<$0.01 mag.  A recent
announcement by Edge et al.\ (2005b) shows they have independently found a
similar period (P=268$\pm$1.4 d) using OGLE and MACHO data. 

The optical data were also examined for shorter periods, looking for
either QPV or NRP.  There do not appear to be any significant periods less
than a day, but there is a weak signal at $\sim$5.3 days, suggesting that
QPV may be present after some outbursts.  This period is present in both
the full data set and when we isolate data taken between the orbital
outbursts.  Its significance is at the 98\% confidence level.  However, as
discussed in Section 2, the power of QPV signals may be weaker in the full
data set since there is not a single, stable period.  In Fig.\ 7 we plot
the $R^*$ light curve folded on the 5.3-day period. 

\subsection{ RX~J0054.9$-$7226 = XTE~J0055$-$724 }

The 58.9s pulsar XTE~J0055$-$724 was discovered by Marshall et al.\
(1998).  It is identified with MACHO 207.16259.23 and the emission line
star MA93\#810.  Its longterm $R$ light curve is shown in the upper panel of
Fig.\ 8.  From a subset of OGLE-II data, Coe \& Orosz (2002) found a peak
in the frequency spectrum corresponding to a period of 14.26 days, but
they say no obvious light curve was present when they folded the data on
this period. 

We have analyzed all of the OGLE-II data as well as the $V$ and $R$
photometry from the MACHO project.  Periodograms for detrended $R$ and $I$
data show power near 15, 20, and 30 days, but these turn out to be aliases
of the true period.  Because of the presence of multiple periods, we also
examined portions of the data in more detail -- that is, divided into
smaller time segments and by MACHO pier orientation.  Each subsection of
data gave somewhat different power for periods near 15, 20, and 30 days.
However, analysis using the PDM method shows the fundamental (orbital)
period is 60.2$\pm$0.8 days, as is seen in both $R^*$ and $I^*$ data in
Fig.\ 8.  The non-sinusoidal shape of this light curve produces no
prominent power in the periodogram at the fundamental period.  This is
similar to the behavior seen in XMMU~J004723.7$-$731226 (discussed above)
where the aliases are simple fractions of the orbital period. 

The folded $R^*$ and $I^*$ light curves can been seen in Fig.\ 9. They
have a peculiar shape, which looks like a sinusoidal variation with a
superimposed outburst near phase zero.  The outburst itself barely rises
above the mean light level.  The light curve resembles that found for
XMMU~J004723.7$-$731226 (see Fig.\ 1), although the `outburst' here is
less pronounced. 

Using RXTE data, Laycock et al.\ (2004) proposed a period of
P$_X$=123$\pm$1 days based on the spacing between four X-ray outbursts. 
We note that this is about double the optically determined period.  We
also searched for short periods that might arise from either QPV or from
NRP, but we found nothing significant. 

\subsection{XMMU~J005517.9$-$723853 = RX~J0055.2$-$7238 }

RX~J0055.2$-$7238 is a 701.6 s X-ray pulsar (Haberl et al.\ 2000; Haberl
et al.\ 2004) which is identified with MACHO 207.16313.35.  Its longterm
light curve is shown in Fig.\ 10.  Clearly there is shortterm variability
around a mean magnitude of $V\sim$16.1.  The slight downward trend in both
$V$ and $R$ was removed before further analysis of the data. 

The detrended photometric data reveal a pronounced period at P=0.28 days,
much too short to be either an orbital period or even the Be star's
rotational period.  The variation is approximately sinusoidal with
amplitudes of $\Delta$$V\sim$0.03 and $\Delta$$R\sim$0.02 mag.  This short
period is most likely due to nonradial pulsations of the type seen in
some single Be stars (e.g. Balona \& James 2002).  This is the second
example (see XTE~J0052$-$725 above) of NRP discovered in this study. 

The periodogram from the $V^*$ data is plotted in Fig.\ 10, where aliases
of the 0.28-day period are also seen.  The same period is found using
$R^*$ data.  The bottom panel shows the $V^*$ and $R^*$ light curves
folded on this very short period.  Porter \& Rivinius (2003; also see
other references therein) in their review of classical Be stars report
that in single Be stars, nonradial pulsations are more common in early
spectral types.  Haberl et al.\ (2004) found that the color of
RX~J0055.2$-$7238 implies the primary star has a very early spectral type
(O9V), making it consistent with the behavior of single Be stars. 

When the MACHO data are subdivided into 4 time segments (A-D), we find
that the period changed with time.  Expanded periodograms for each segment
are shown in Fig.\ 11.  In the lower panel we plot period versus time,
showing that the period decreased from 0.28484 days to 0.28470 days at a
mean rate of $-2.6$ s yr$^{-1}$.  NRP in Be stars are well known to
exhibit changes in periods, amplitudes, and light curve shapes similar to
those observed here (e.g. Balona, Sterken, \& Manfroid 1991). 

Analysis of the $R^*$ data taken before and after removal of the 0.28-day
variation shows weak power at P$\sim$412$\pm$4 days in both the
periodogram and the PDM.  The power spectrum in Fig.\ 12 shows the 412-day
peak has a confidence level below 90\%.  If real, we assume this is the
orbital period, which would be in reasonable agreement with the Corbet
(1984) P(pulse)/P(orbit) relation.  The 412-day $R^*$ light curve shows
$\sim$0.01 mag variation (bottom panel of Fig.\ 12), similar to the
behavior found in some other Be/X-ray systems.  If the orbit is not very
eccentric it is possible that there is little interaction between the
neutron star and the Be star's equatorial disk, resulting in only a very
small change in system brightness. 

\subsection{ Reanalysis of RX~J0050.7$-$7316 = AX~J0051$-$733 = DZ~Tuc } 

The discovery of this 323 s pulsar was announced by Yokogawa \& Koyama
(1998b) and discussed in more detail by Imanishi et al.\ (1999).  The
optical counterpart was identified as a Be star by Cowley et al.\ (1997)
in their study of X-ray sources in the Magellanic Clouds, and it is
coincident with MA93\#387.  The colors ($B-V=-0.03$ and $U-B=-0.95$)
indicate that the primary is a very early B or late O star. 

Using OGLE-II data, Coe \& Orosz (2000) found a period of 0.708 days, as
had been suggested by Cook (1998), which they interpreted as half the
orbital period in a close binary.  Further analysis using a subset of
MACHO data as well as OGLE-II data was carried out by Coe et al.\ (2002)
showing that both the period and amplitude change with time.  If the
period change is linear, it amounts to $-$13.5 s yr$^{-1}$, with the $R$
amplitude changing by $\sim$40\%.  In spite of the extreme nature of these
parameters, they concluded that the system was likely to contain a very
close binary with P$_{orb}$$\sim$1.4 days. 

We have reanalyzed the photometry for this system, now using the complete
MACHO data set which adds more than 3 years of data.  The longterm $V$
light curve is shown in Fig.\ 13.  The amplitude clearly changes with
time, with the least variability occurring around MJD 50000.  We also
found that there are slightly different mean magnitude levels in the MACHO
data depending on whether the telescope was east or west of the pier.
Thus, we flattened the data from these two configurations separately and
then combined them to search for periodicities. 

Our period analysis was done first using the entire data set and then
subdividing the data into time segments of $\sim$200 days, some of which
are indicated on Fig.\ 13.  In the middle panel of Fig.\ 13 we show the
power spectrum for several sample segments, demonstrating that the period
changes.  Light curves, folded on the $\sim$0.7-day period, are shown for
two of these segments to emphasize how much the amplitude varies.  Within
a given time segment, there is little or no color variation through the
0.7-day period. 

In Fig.\ 14 we plot the photometric period and amplitude versus date,
using only segments which contain more than 30 observations.  The
character of the variation, with both period and amplitude changing with
time, suggests that RX~J0050.7$-$7316 is another nonradial pulsator, as
was found above for XTE~J0052$-$725 and XMMU~J005517.9$-$723853.  Some
single Be stars are known to show pulsation periods near 0.7 days (e.g.
Balona et al.\ 1991, Balona 1992).  The 1.4-day binary interpretation
always required special circumstances, such as rapid mass transfer, to
explain all of the observations.  Instead, interpreting the variations as
due to NRP seems more probable to us, since two other systems with very
short periods have been found.  In addition, the early spectral type of
this Be star suggests it might be likely to show NRP, as discussed above. 

In order to search for a longer binary period we removed the 0.7-day
variations and examined the residuals.  Although we searched up to P=1000
days, there was no power strong enough at any frequency to be convincing
evidence of orbital outbursts. 

Imanishi et al.\ (1999) suggested their X-ray data were consistent with a
185-day period, but they stressed that further monitoring would be needed.
Laycock et al.\ (2004) proposed a possible 108$\pm$18 day X-ray period. 
We find nothing in the optical data to support either of these suggested
periods.  Perhaps the orbit is fairly circular so that the neutron star
does not come close enough to the Be star to cause an optical outburst,
hence leaving the primary to behave like an isolated star. 

\section{ Discussion and Summary}

In this paper we have investigated the longterm light curves of 7 Be/X-ray
pulsars in the Small Magellanic Cloud.  We have shown that multiple types
of optical variability are present including longterm irregular changes,
orbital outbursts, quasi-periodic variations of the disk, and nonradial
pulsations of the Be star. 

For 5 of the systems studied, we have found low amplitude periodic
outbursts which we interpret as being due to the interaction of the
neutron star with the Be star's disk once during each orbit, probably near
periastron passage.  Corbet (1984) and others have shown that the X-ray
pulsation period and orbital period are correlated in Be/X-ray binaries.
In Fig.\ 15 we plot a ``Corbet diagram", including the systems in this
study as well as other Be/X-ray binaries (both in the Galaxy and the
Magellanic Clouds) with published orbital periods. 

For two systems (RX~J0050.7$-$7316 and XTE~J0052$-$725) we did not find
any signature of the orbital period in the light curve.  However, for both
of these systems we discovered that NRP are present.  These nonradial
pulsators may either be in sufficiently wide orbits that the Be star
behaves like a single star or the system could have a low eccentricity so
there is little interaction between the stars.  Although such pulsations
are known to be present in some single Be stars, this is the first time
they have been identified in Be/X-ray binaries.  In addition, we found
that XMMU~J005517.7$-$723853 shows both strong NRP and a small outburst
every $\sim$412 days (probably the orbital period). 

RX~J0049.1$-$7250 and CXOU~J005455.6$-$724510 show both orbital outbursts
and quasi-periodic variations of the type previously recognized by
Schmidtke et al.\ (2004).  The strong signature of the 33-day orbital
period in RX~J0049.1$-$7250 suggests that its orbit may be significantly
eccentric. 

In summary, we have discovered that nonradial pulsations are present in
some Be/X-ray systems.  We have also determined orbital periods for 5
systems and found 2 which show quasi-periodic variations.  The Be/X-ray
binaries may be divided into at least two groups.  The systems in which
the Be star has NRP either show no orbital outbursts or very weak ones,
suggesting that they are in wide or nearly circular orbits (or both) with
little interaction between the two stars.  The systems in which the
orbital interaction is strong (outbursts present) often exhibit QPV
following the outbursts.  None of the studied binaries with QPV also shows
nonradial pulsations.  The remaining two systems in which the orbital
signature is weak show neither QPV nor NRP.  

To further investigate these types of behavior, we reexamined the
photometry of the 4 Be/X-ray binaries studied previously by Schmidtke et
al.\ (2004) which display both orbital and QPV variations.  We searched
for very short periods which would indicated the presence of nonradial
pulsation, but none was found.  Hence these binaries behave like the ones
in the present study, with QPV and NRP not occurring in the same system.

\acknowledgments
We thank Robin Corbet, Malcomb Coe, and Jose Galache for helpful
information about some of these systems.  We also thank the anonymous
referee for useful suggestions which improved the paper.

\clearpage

\begin{deluxetable}{llcc}
\tablenum{1}
\tablecaption{Small Magellanic Cloud Be/Neutron-Star Systems Studied}
\tablehead{  
\colhead{System Name} &
\colhead{R.A.(2000)\tablenotemark{a}} &
\colhead{Dec.(2000)\tablenotemark{a}} &
\colhead{$<R>$\tablenotemark{b} } \\
\colhead{~~~~~MACHO \#} &
\colhead{~~~~~OGLE-II \#} &
\colhead{~~~~~MA93 \#\tablenotemark{c} } &
\colhead{~~~~~~~~${V-R}$ } 
}
\startdata
XMMUJ004723.7$-$731226  & 00:47:23.7 & $-$73:12:26.9 & 16.2  \nl
~~~~~212.15792.77 & ~~~~~004723.37$-$731226.9 & ~~~~~172 & ~~~~~~~0.05 \nl
RXJ0049.1$-$7250 & 00:49:03.3  & $-$72:50:52 & 16.9  \nl
~~~~~208.15911.93 & ~~~~~004903.35-725052.1 & ~~~~~- - - & ~~~~~~~0.15 \nl
RXJ0050.7$-$7316 & 00:50:44.7  & $-$73:16:05 & 15.5  \nl
~~~~~212.16019.30 & ~~~~~005044.71$-$731605.0 & ~~~~~387 & ~~~~~~~0.15 \nl
XTEJ0052$-$725 & 00:52:09.1  & $-$72:38:03 & 15.0  \nl
~~~~~208.16085.24 & ~~~~~005208.95$-$723802.9 & ~~~~~- - - & ~~~~~~~0.17 \nl
CXOUJ005455.6$-$724510 & 00:54:55.9 & $-$72:45:10.5 & 15.0 \nl
~~~~~207.16254.16 & ~~~~~- - - & ~~~~~809 & ~~~~~~~0.12 \nl
RXJ0054.9$-$7226 & 00:54:56.2 & $-$72:26:47.6  & 15.4  \nl
~~~~~207.16259.23 & ~~~~~005456.17-722647.6 & ~~~~~810 & ~~~~~~~0.14 \nl
XMMUJ005517.9$-$723853  & 00:55:17.9 & $-$72:38:53 & 15.9 \nl
~~~~~207.16313.35 & ~~~~~- - - & ~~~~~- - - & ~~~~~~~0.21 \nl
\enddata

\tablenotetext{a}{Coordinates are the MACHO position of optical star. }

\tablenotetext{b}{Approximate mean $R$ magnitude. }

\tablenotetext{c}{from catalogue of H$\alpha$ emission line stars
in the SMC by Meyssonnier \& Azzopardi (1993). } 

\end{deluxetable}

\newpage

\begin{deluxetable}{lcccl}
\tablenum{2}
\tablecaption{Photometric Periods for 
Small Magellanic Cloud Be/Neutron-Star Systems}
\tablehead{  
\colhead{} &
\colhead{} & 
\multicolumn{2}{c}{Photometric Periods} & 
\colhead{Periods in} \\
\colhead{System} & 
\colhead{P$_{pulsar}$} & 
\colhead{P$_{orb}$} &
\colhead{Other P\tablenotemark{a}} &
\colhead{Literature} \\  
\colhead{} &
\colhead{(sec)} & 
\colhead{(days)} & 
\colhead{(days)} & 
\colhead{}
}

\startdata
XMMU~J004723.7$-$731226  & 263.6    & 49.1$\pm0.2$  & - - -  & 
Edge et al.\tablenotemark{b}    \nl
RX~J0049.1$-$7250       &  74.7    & 33.4$\pm0.4$  & QPV$\sim$2.4 & 
Galache\tablenotemark{c}   \nl
RX~J0050.7$-$7316       & 323      & - - -    & NRP$\sim$0.706  & 
Coe \& Orosz\tablenotemark{d}  \nl
XTE~J0052$-$725          &  82.4    & - - -    & NRP$\sim$1.328 \nl
CXOU~J005455.6$-$724510 & 503.5 & 273$\pm6$  & QPV$\sim$5.3 & Edge et 
al.\tablenotemark{e} \nl
RX~J0054.9$-$7226        &  58.9    & 60.2$\pm0.8$ & - - - & 
Laycock et al.\tablenotemark{f}     \nl
XMMU~J005517.9$-$723853  & 701.6    & (412$\pm4$)\tablenotemark{g} & 
NRP$\sim$0.285  \nl

\enddata

\tablenotetext{a}{QPV = quasi-periodic variations of Be
star's equatorial disk (see text and Schmidtke et al.\ 2004); NRP =
nonradial pulsations of Be star; P$_X$ = period from X-ray outbursts. } 

\tablenotetext{b}{Edge et al.\ (2005a) give P$_I$=48.8$\pm$0.6 days from
OGLE data (see text for discussion). } 

\tablenotetext{c}{Galache (private communication 2005) reports
P$_X\sim65.1$ days, while Laycock et al.\ (2004) suggest P$_X\sim642\pm59$
days.  Galache's period is approximately twice the optically determined
one, suggesting X-ray outbursts may not occur in each orbit or not all
X-ray outbursts have been observed. } 

\tablenotetext{d}{Coe \& Orosz (2000) discuss a binary interpretation with
P$_{orb}$=1.4 days.  Laycock et al.\ (2004) give P$_X\sim108\pm$18 days;
Imanishi et al.\ (1999) find weak evidence for P$_X\sim$185 days.} 

\tablenotetext{e}{Edge et al.\ (2005b) independently find
P$_{orb}$$\sim$268$\pm$1.4 days.} 

\tablenotetext{f}{Laycock et al.\ (2004) suggest P$_X\sim123\pm$1 days,
which may be double the optical period.} 

\tablenotetext{g}{This value is quite uncertain; see text for discussion.}

\end{deluxetable}

\clearpage

\begin{figure}
%fig 1 
\caption{(top) Longterm $V$ light curve of XMMU~J004723.7$-$731226.
~(second) Upper part of this panel shows the variance calculated using PDM
analysis for $V^*$ data after MJD 49350; the scale is on the right.  The
lower part of this panel displays the periodogram for the same data.  Note
that the orbital period (P=49.1 d) is not seen in $V^*$ data, although the
alias at 9.8 d may be marginally present in the periodogram.  The 99\%
confidence level for peaks in the periodogram is shown by the horizontal
dotted line.  ~(third) Similar to middle panel, with the upper portion
showing the variance from the PDM analysis of $I^*$ data and the lower
curve being the periodogram for the same data.  The 9.8 and 12.3 day
aliases show strongly in the periodogram, but the PDM reveals 49.1 days to
be the fundamental period.  The peaks at low frequency ($<$0.01 d$^{-1}$)
are caused by the spacing of the OGLE data and are not present in the
MACHO data.  ~(bottom) Binned $V^*$ and $I^*$ light curves of
XMMU~J004723.7$-$731226 folded on P=49.09 days with phase zero at MJD
50004.4.  Only the $I^*$ curve shows the periodic outbursts. } 
\end{figure}

\begin{figure}
%fig 2 
\caption{(top) Longterm $R$ light curve of RX~J0049.1$-$7250.  The data
were detrended in two segments, as marked. ~(second) Periodogram using
$R^*$ data from combined segments A and B.  There is strong power at
P=33.4 days, which is likely to be the orbital period.  The multiple
frequencies with power near 2.4 days are from quasi-periodic variations
which follow each outburst.  The dotted horizonal line shows the 99\%
confidence level for these periods. The peaks at very low frequencies are
related to the length of the data sample. ~(third) $R^*$ light curve from
Seg A and Seg B, folded on the 33.4-day period.  ~(bottom) $R^*$ and $V^*$
magnitudes from Seg A folded on the QPV period of 2.4 days.  Note that
both colors show approximately the same amplitude. } 
\end{figure}

\begin{figure}
%fig 3 
\caption{(top) Longterm $I$ light curve for RX~J0049.1$-$7250 from OGLE-II
data.  Note the unusually strong outburst near MJD 50760. ~(middle)
Periodogram from $I^*$ data, excluding the super-outburst, showing
pronounced power at the 33-day orbital period as well as the QPV at
$\sim$2.4 days.  As in previous figures, the 99\% confidence level for
these periods is shown by the dotted line.  ~(bottom) $I^*$ light curve
folded on the 33.4-day orbital period with phase zero at MJD 50031.0.
Filled symbols show the binned data, excluding the super-outburst; open
symbols show the individual observations during the super-outburst near
MJD 50760. } 
\end{figure}

\begin{figure}
%fig 4 
\caption {(top) The longterm $R$ (upper) and $I$ (lower) light curves of
XTE~J0052$-$725.  The scale for the $I$ data is on the right.  The data
were analyzed in time segments (A-E), as marked.  ~(middle) Periodogram
from $R^*$ data in combined Seg A, B, and C, showing strong power at
frequency F, which corresponds to P=1.328 days.  Aliases of this period
are also indicated. ~(bottom) Periodograms from Seg A, B, C, and D,
expanded around the primary peak, showing how the power of the nonradial
pulsations decreased with time, although the period itself remained quite
constant. } 
\end{figure}

\begin{figure}
%fig 5 
\caption {(top) Binned $R^*$ and $V^*$ light curves of XTE~J0052$-$725
from Seg A folded on P=1.328 days.  ~(second) Similar to top panel for Seg
B.  ~(third) Similar to top panel for Seg C.  One can see how the
amplitude of the variation decreased with time.  There doesn't appear to
be any color term.  ~(bottom) Individual $I$ points (not binned) from Seg
E folded on the mean period from Seg A and B.  The amplitude in this
segment is similar to that seen in Seg A, but the phasing shifted when the
pulsations resumed. } 
\end{figure}

\begin{figure}
%fig 6 
\caption{(top) Longterm $R$ light curve of CXOU~J005455.6$-$724510.
~(middle) Upper part of this panel shows the variance calculated using the
PDM method for $R^*$ data; the scale is on the right.  The lower part of
this panel displays the periodogram for the same data.  The 99\%
confidence level is shown by the dotted line.  The peak closest to 0.002
d$^{-1}$ is not present in the PDM and is probably spurious.  ~(bottom)
$R^*$ light curve folded on P=273 days with phase zero at MJD 50000, where
a $\sim$0.02 mag outburst occurs.  The data have been binned by phase. } 
\end{figure}

\begin{figure}
%fig 7 
\caption{(top) Periodogram of $R^*$ data of CXOU~J005455.6$-$724510
showing a modest peak near 5.3 days.  The 99\% confidence level is shown
by the dotted horizonal line.  ~(bottom) $R^*$ data folded on P=5.29 days
and binned by phase. } 
\end{figure}

\begin{figure} 
%fig 8 
\caption{(top) Longterm $R$ light curve of RX~J0054.9$-$7226. ~(middle)
The upper part of this panel shows the variance calculated using the PDM
method for $R^*$ data from segment A; the scale is on the right.  The
lower part of this panel displays the periodogram for the same data.
Dashed vertical lines show the 60-day orbital period and its aliases,
marked in days.  The dotted horizontal line shows the 99\% confidence
level of the periodogram.  ~(bottom) Similar to the middle panel, but for
$I^*$ data. } 
\end{figure} 

\begin{figure} 
%fig 9 
\caption{(top) $R^*$ light curve folded on P=60.2 days with phase zero at
MJD 50045.0.  Individual points are shown by crosses and binned data by
filled squares. ~(bottom) $I^*$ light curve of RX~J0054.9$-$7226 folded on
the same period, but with only individual observations shown. } 
\end{figure} 

\begin{figure}
%fig 10 
\caption{(top) Longterm $V$ light curve of XMMU~J005517.9$-$723853.  The
dashed vertical lines separate the segments (A to D) which were studied
individually. ~(middle) Periodogram of all segments of $V^*$ data showing
strong power at P=0.28 days.  This period is interpreted as due to
nonradial pulsations (NRP) of the Be star. ~(bottom) $V^*$ and $R^*$
magnitudes from all segments folded on P=0.28 days.  Note the slightly
greater amplitude in $V^*$. } 
\end{figure}

\begin{figure}
%fig 11 
\caption{(top) Expanded periodogram for XMMU~J005517.9$-$723853 in $V^*$
showing the change in period from the four time segments (A-D) marked in
the previous figure. ~(bottom) Period versus time for all time segments,
showing a period change of $-2.6$ s yr$^{-1}$ in both MACHO colors. } 
\end{figure}

\begin{figure}
%fig 12 
\caption{(top) Periodogram from $R^*$ data for XMMU~J005517.9$-$723853,
showing a possible period of 412 days, but with a confidence level of
$<$90\%.  ~(bottom) Binned $R^*$ data plotted the 412-day period, with
phase zero at MJD 50102. } 
\end{figure}

\begin{figure}
%fig 13 
\caption{(top) Longterm $V$ light curve of RX~J0050.7$-$7316 identifying
some of the time segments which were analyzed individually. ~(middle)
Periodograms derived from the segments A-D (as marked in the upper panel),
showing the changing period.  ~(bottom) $V$ light curve from Seg B and C,
folded on the 0.7-day period, to show how much the amplitude changes. 
These light curves are folded on the period unique to their own time
segment. } 
\end{figure}

\begin{figure}
%fig 14 
\caption{(top) Period versus time for multiple time segments in $V$, $R$,
and $I$ of RX~J0050.7$-$7316.  Only segments with more than 30 data points
were used for the period analysis.  The change in period could be
interpreted either as a linear decrease or as an approximately constant
period until MJD 50000, followed by a sudden jump to a shorter period
after that date.  ~(bottom) Amplitude versus time for the same segments as
shown in the top panel.  Note that the amplitude decreased until MJD 50000
and then returned to the original level. } 
\end{figure}

\begin{figure}
%fig 15
\caption{``Corbet diagram" relating X-ray pulsar periods (in seconds) and
orbital periods (in days).  The filled squares are systems in this study.
The other values are taken from the literature, where a variety of
techniques were used to find the orbital periods (X-ray outbursts, optical
photometry, spectroscopy).  } 
\end{figure}

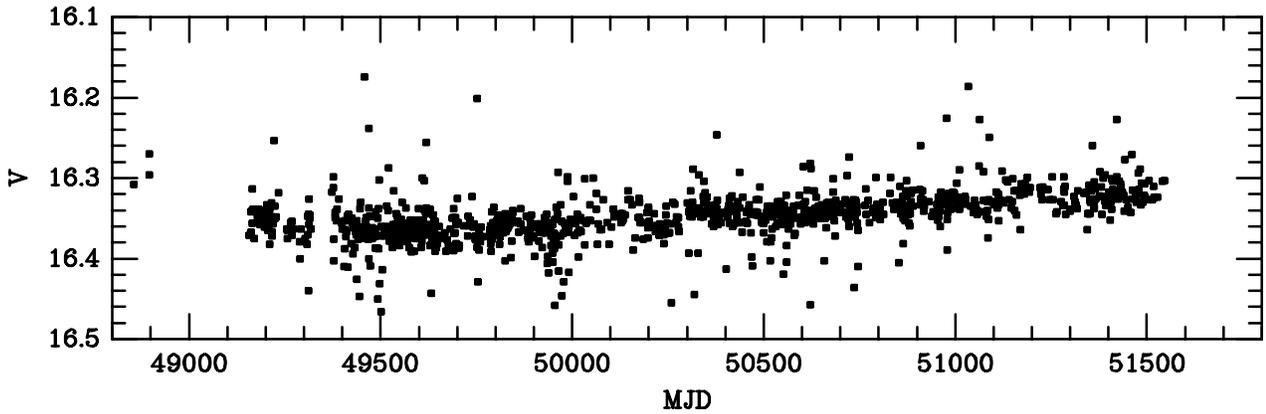
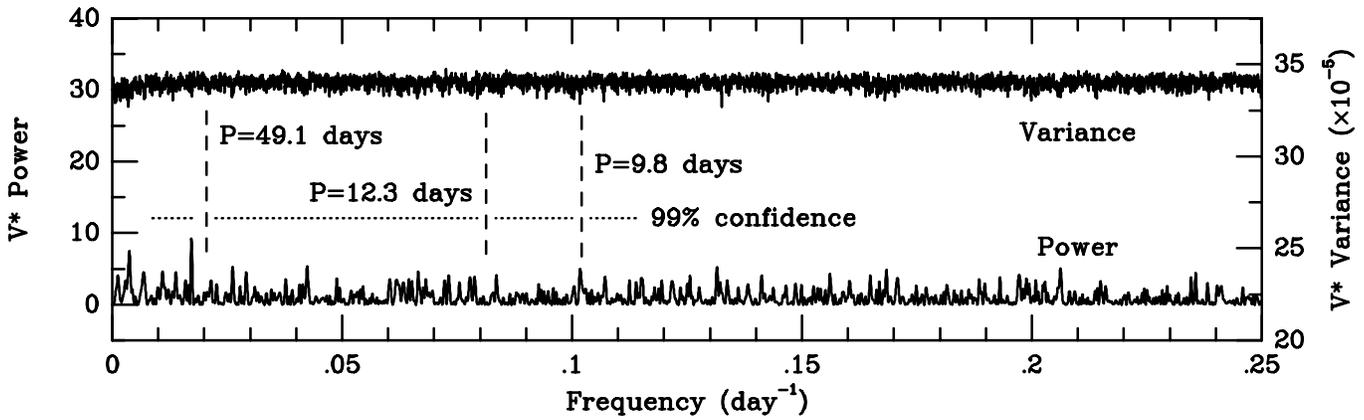
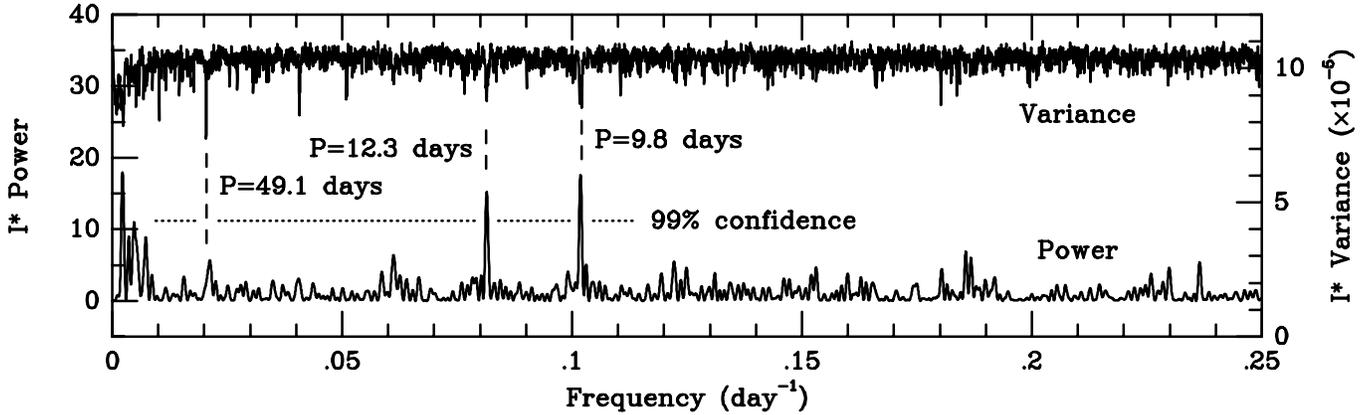
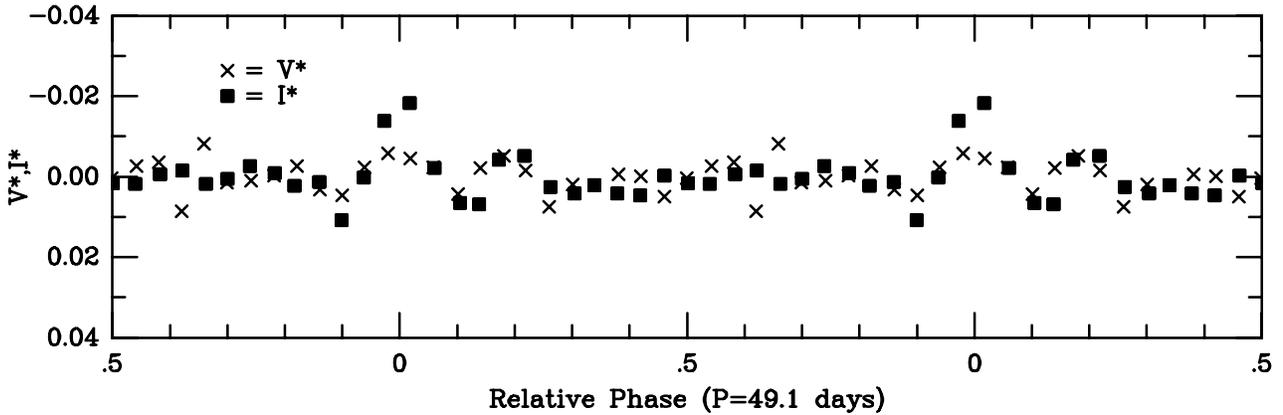
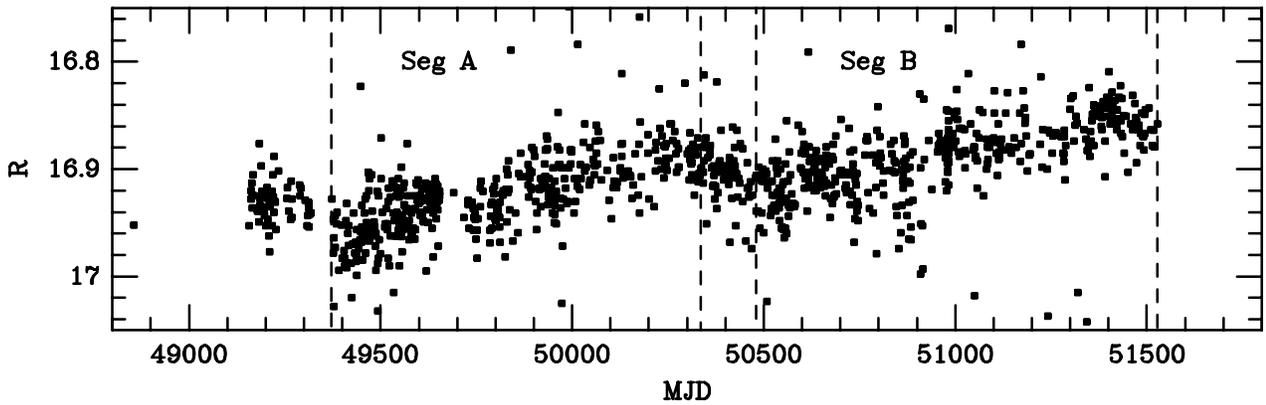
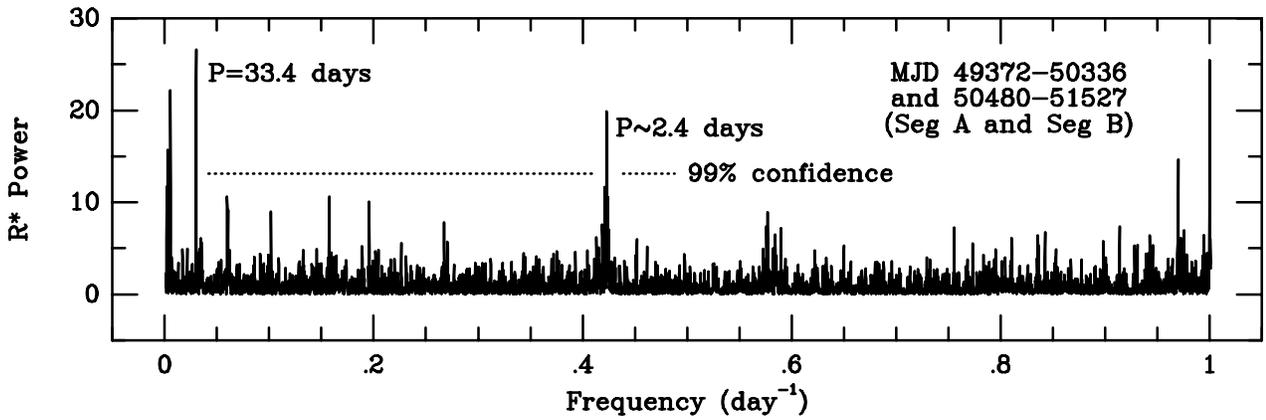
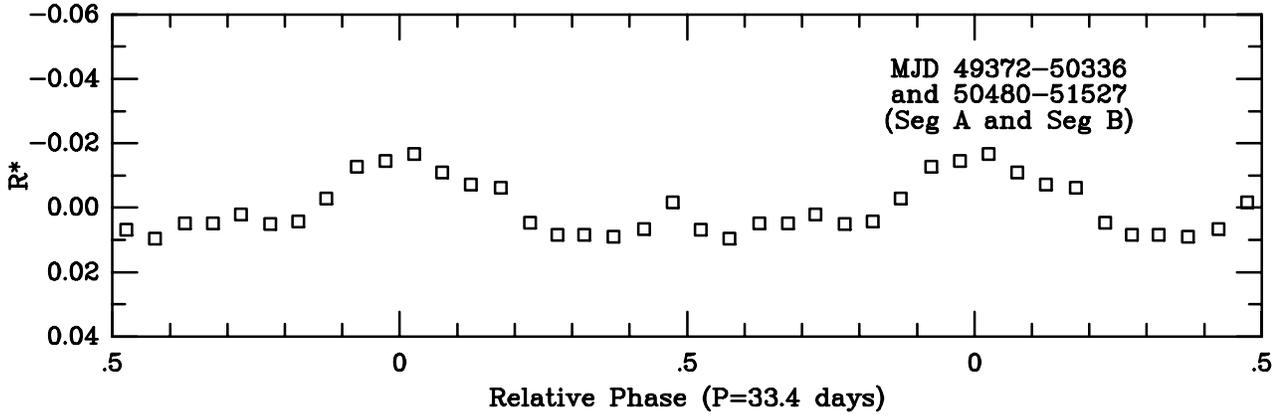
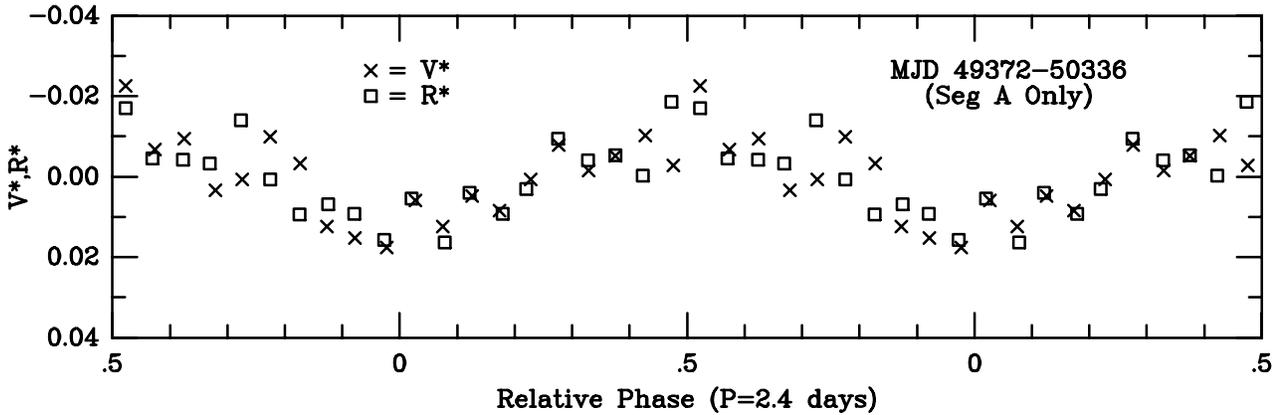
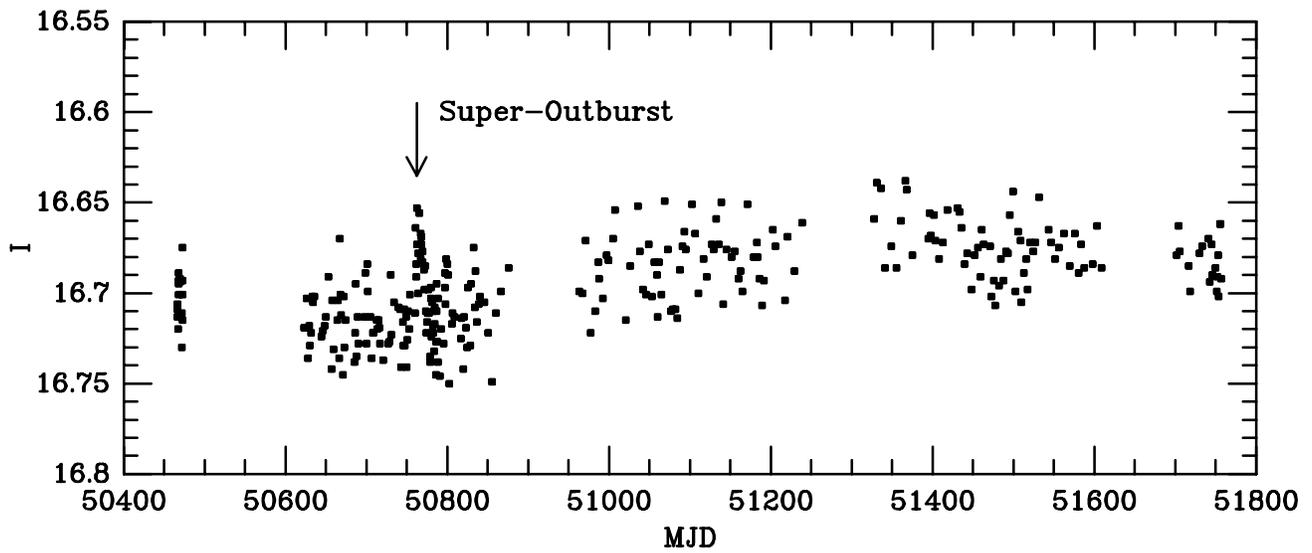
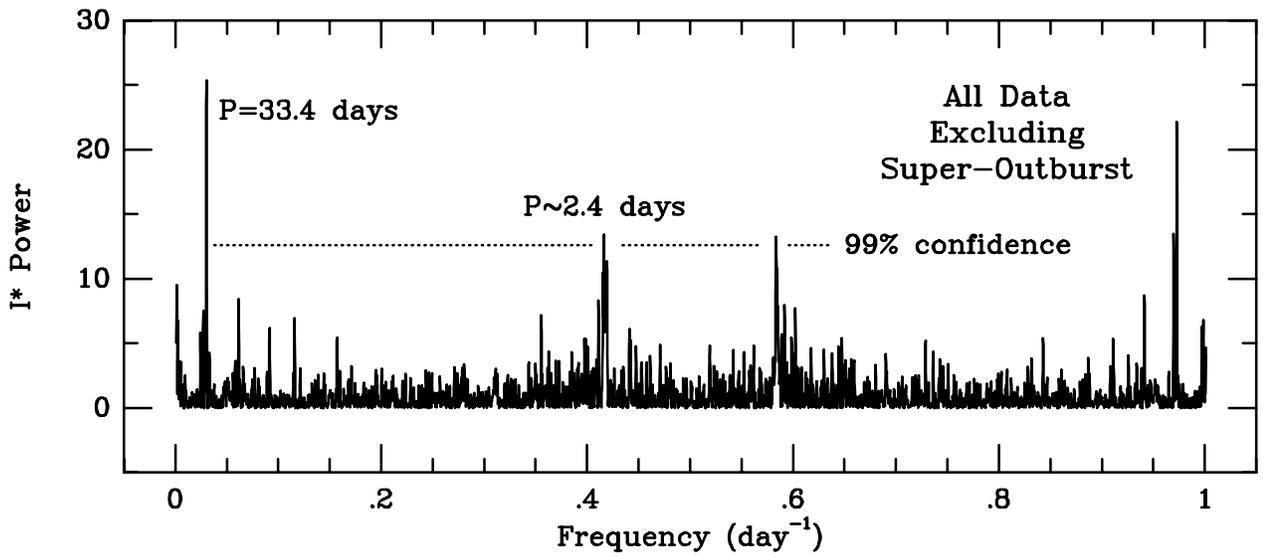
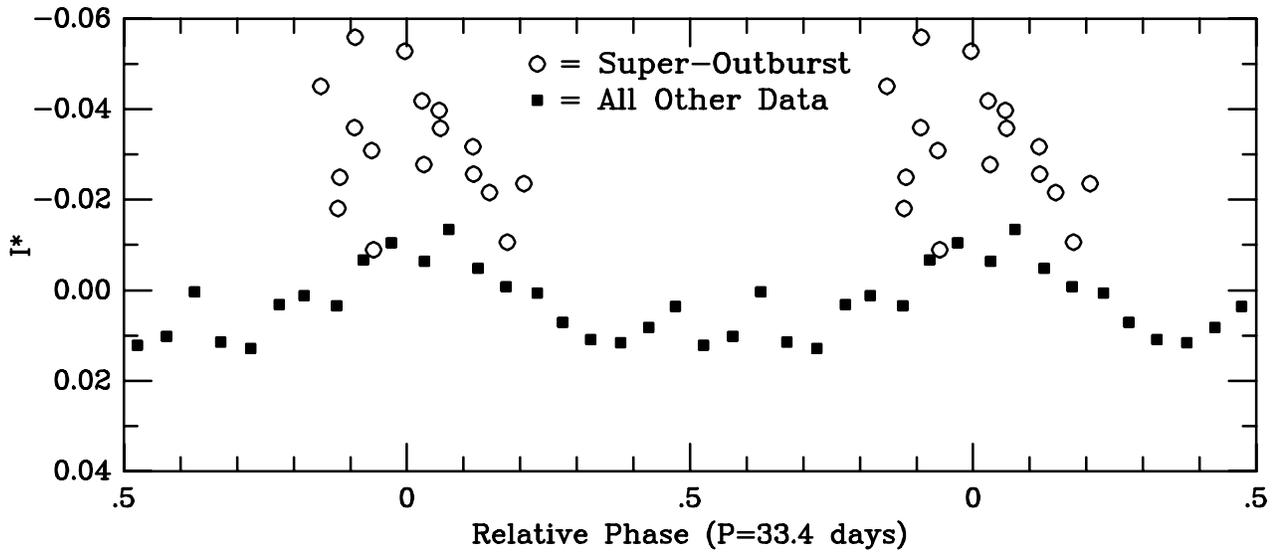
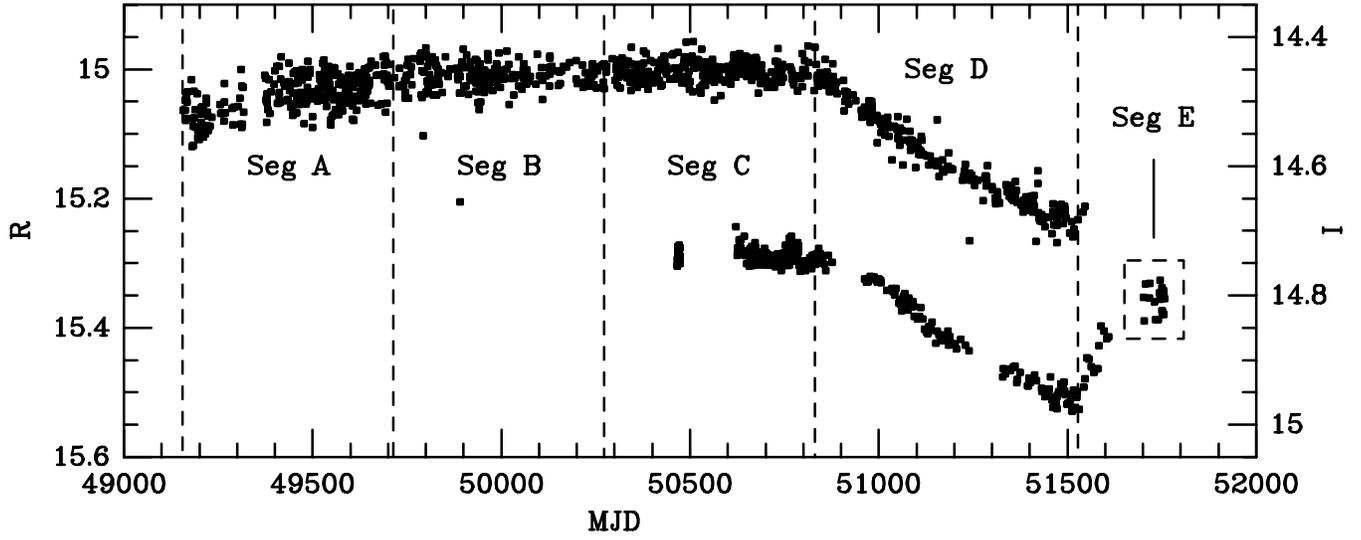
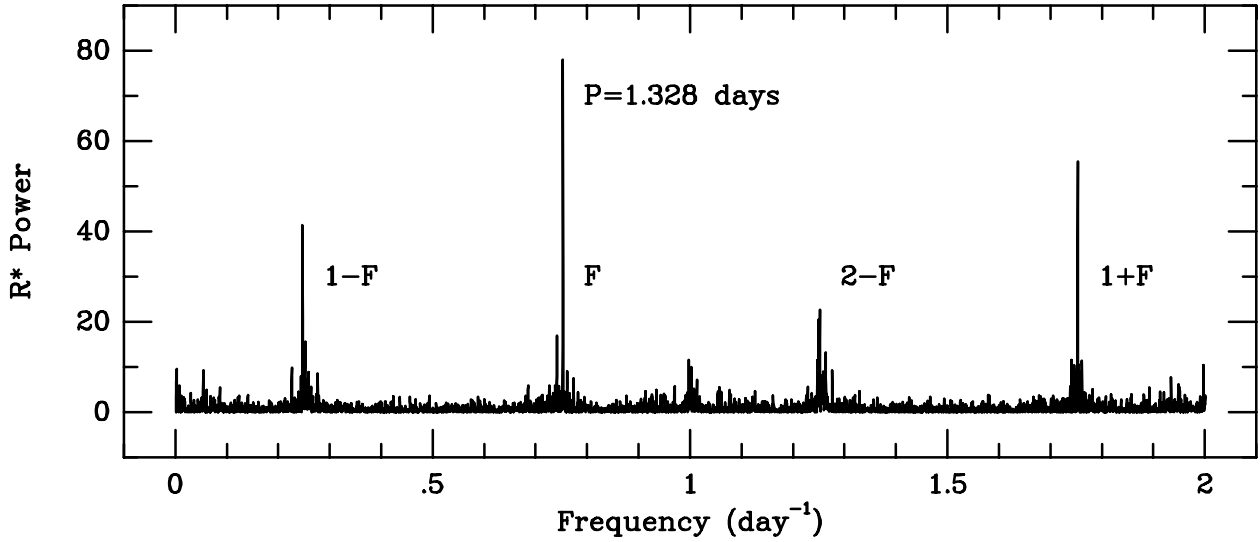
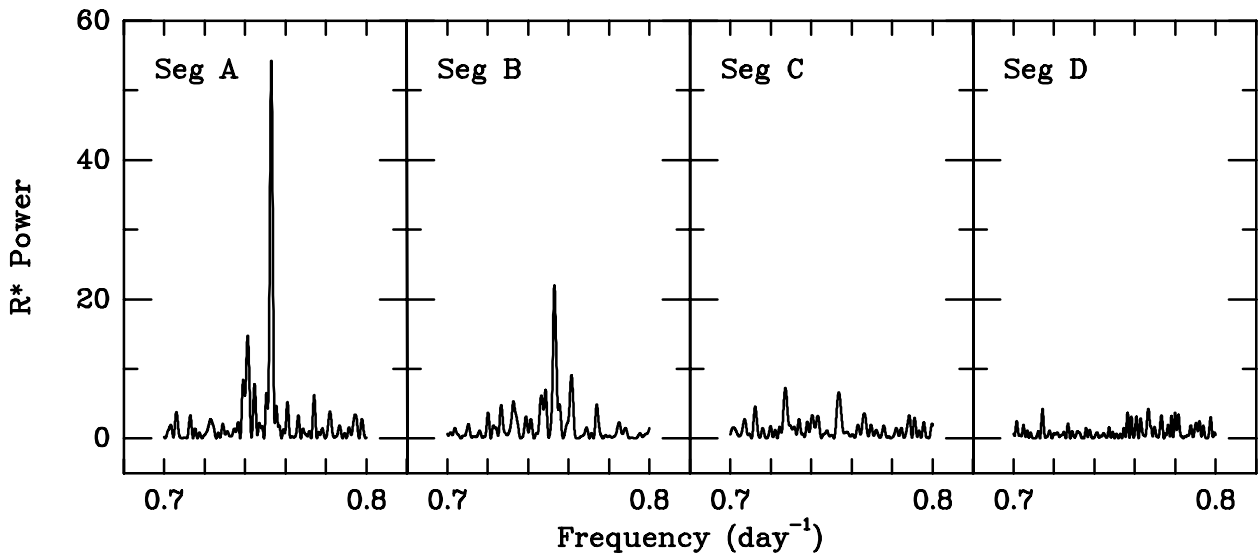
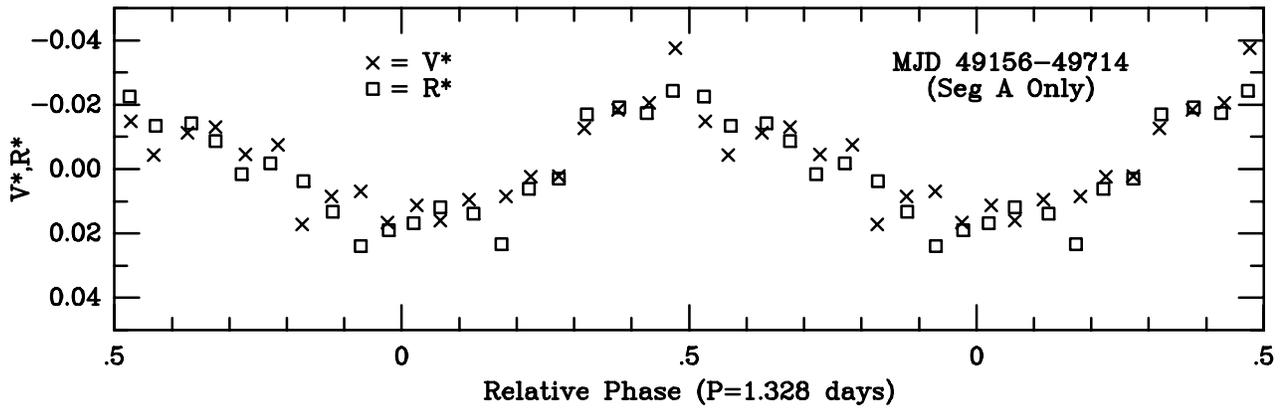
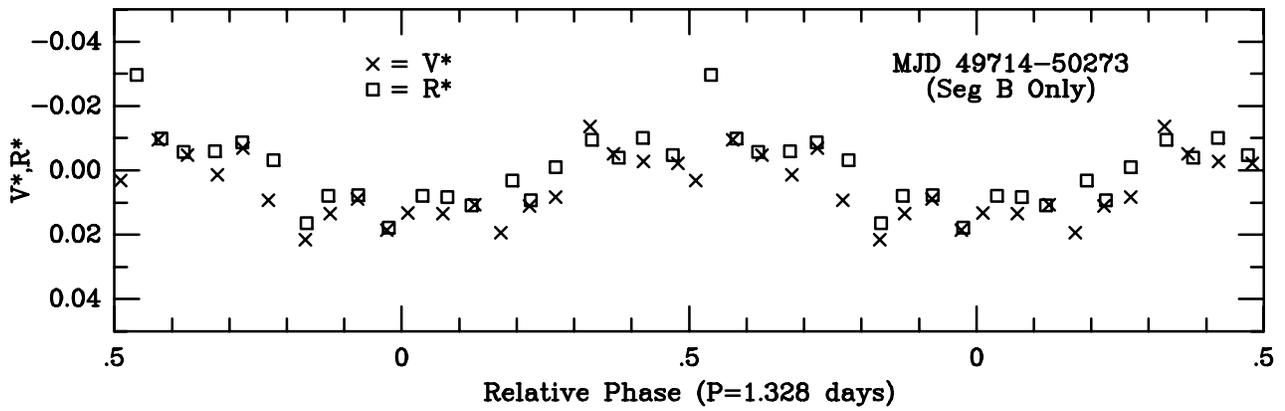
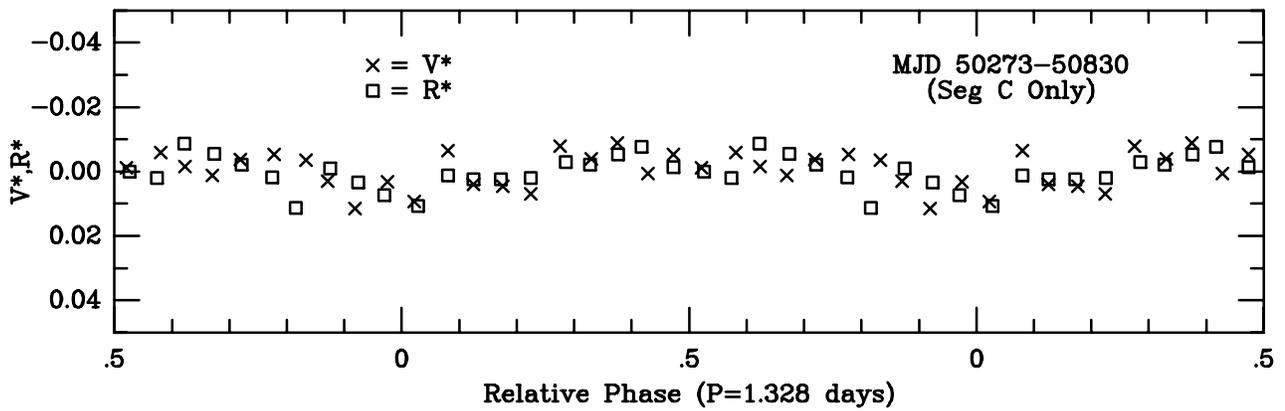
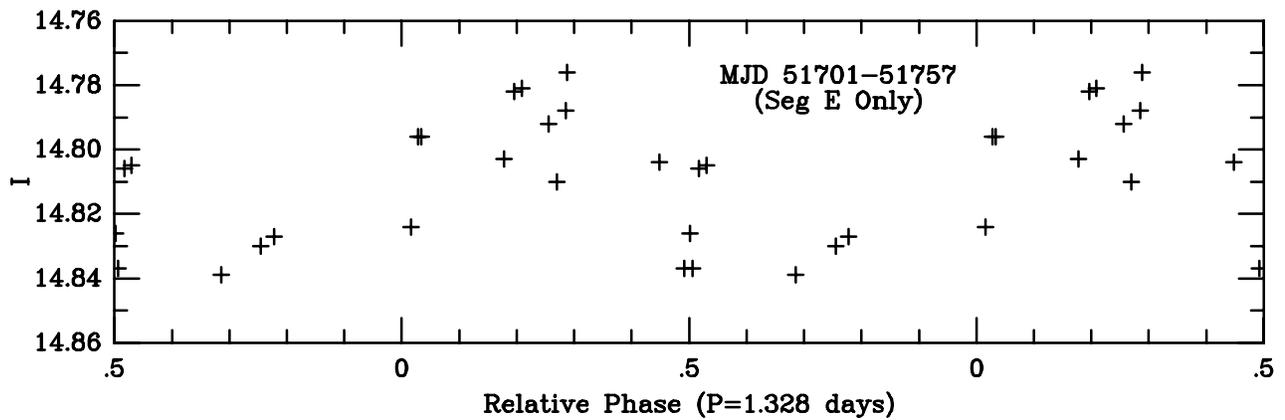
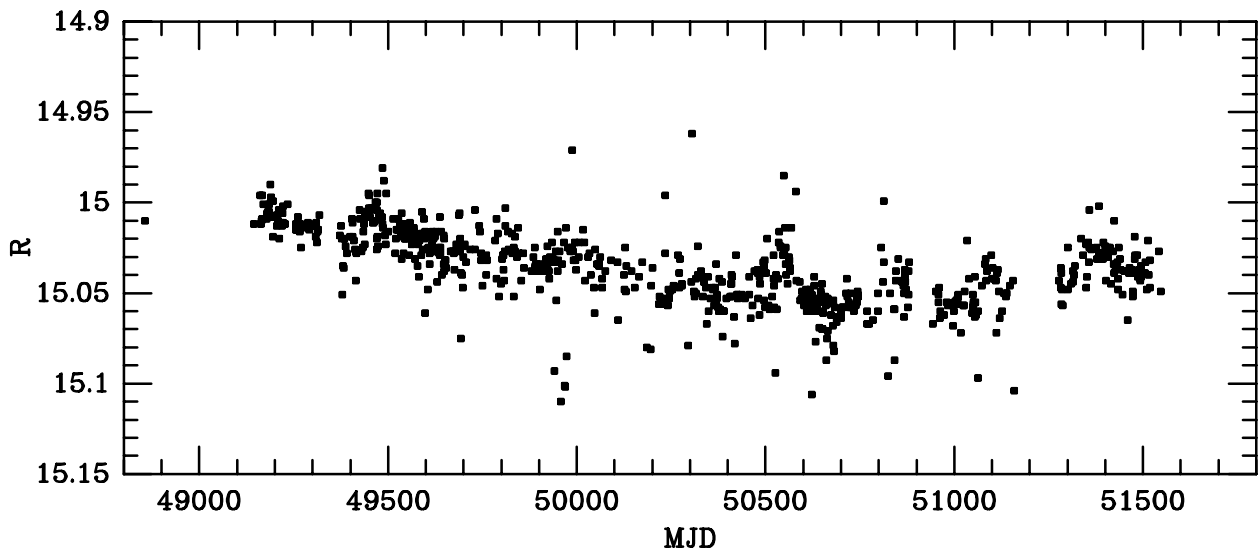
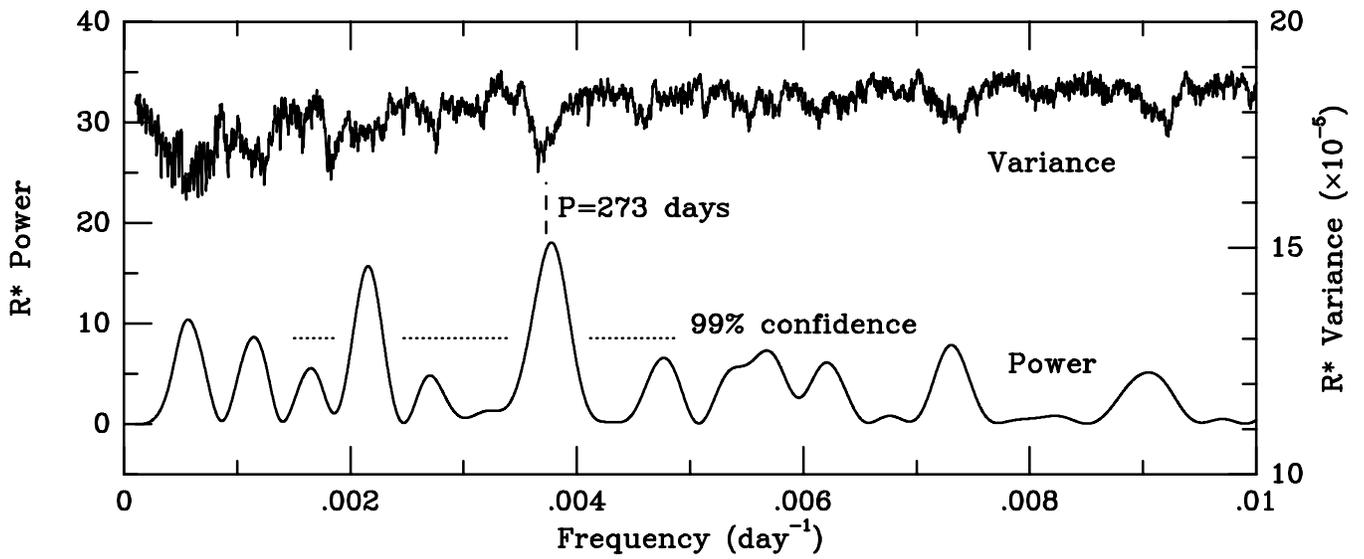
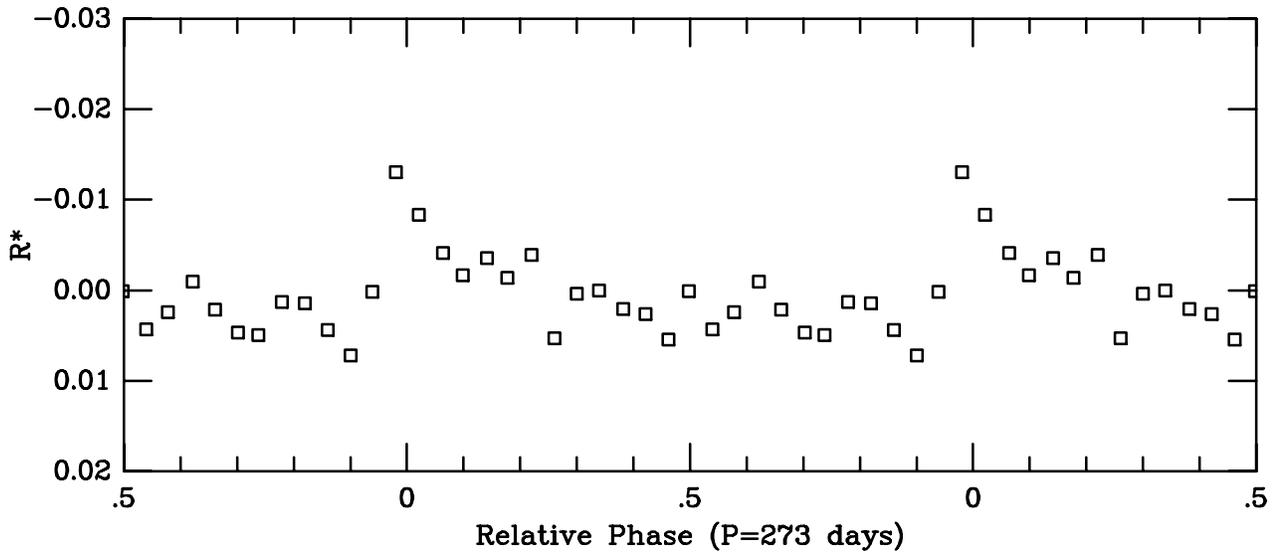
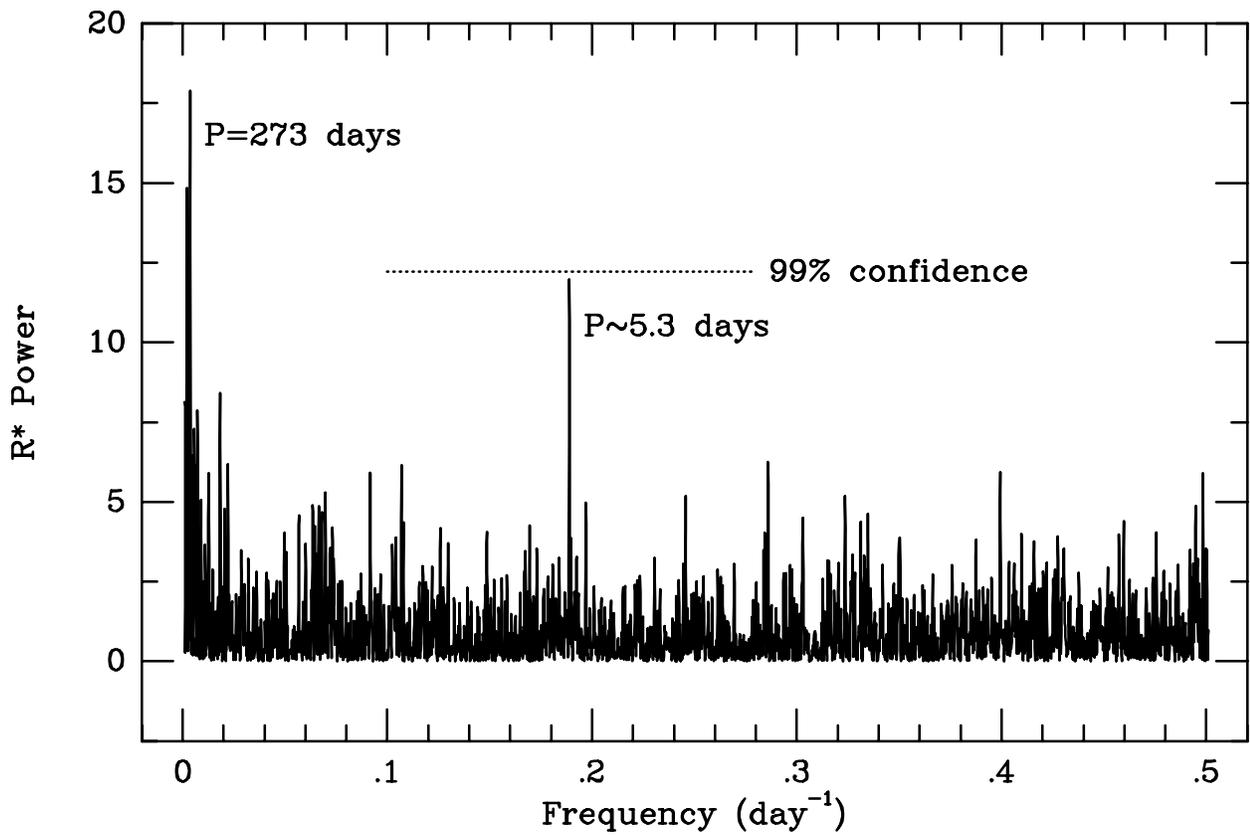
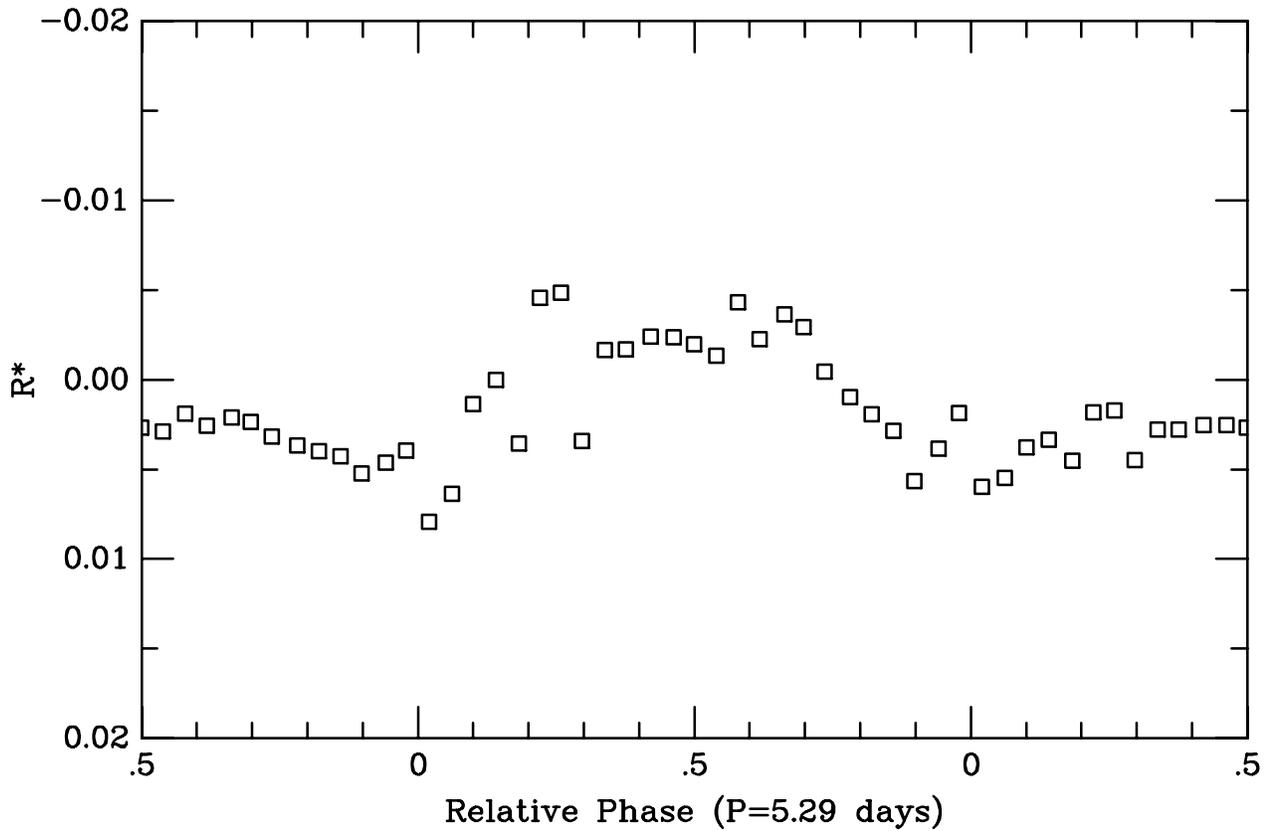
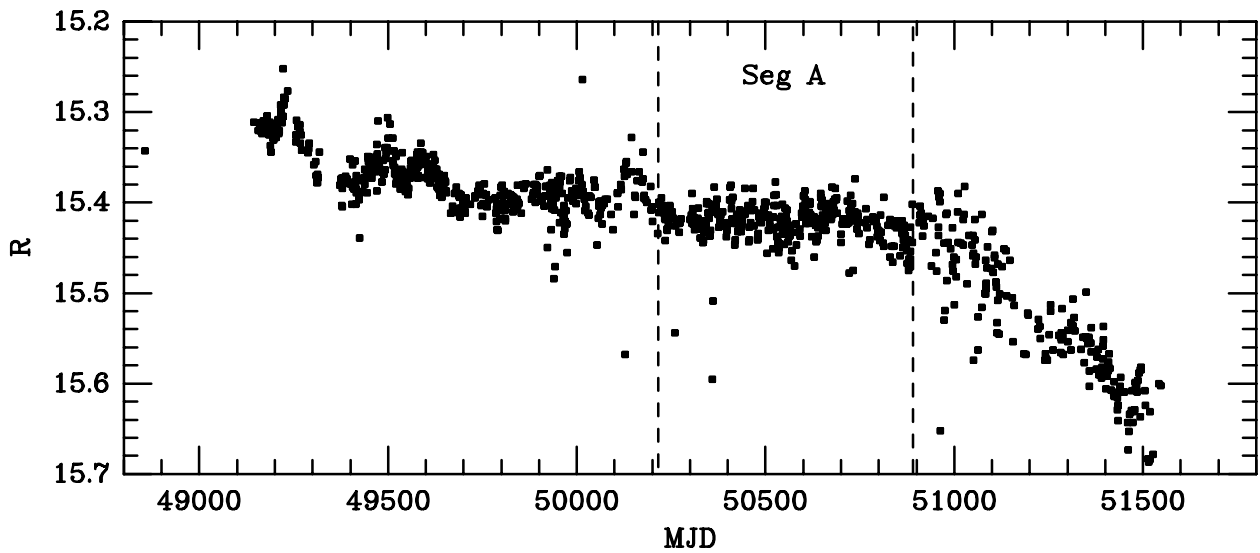
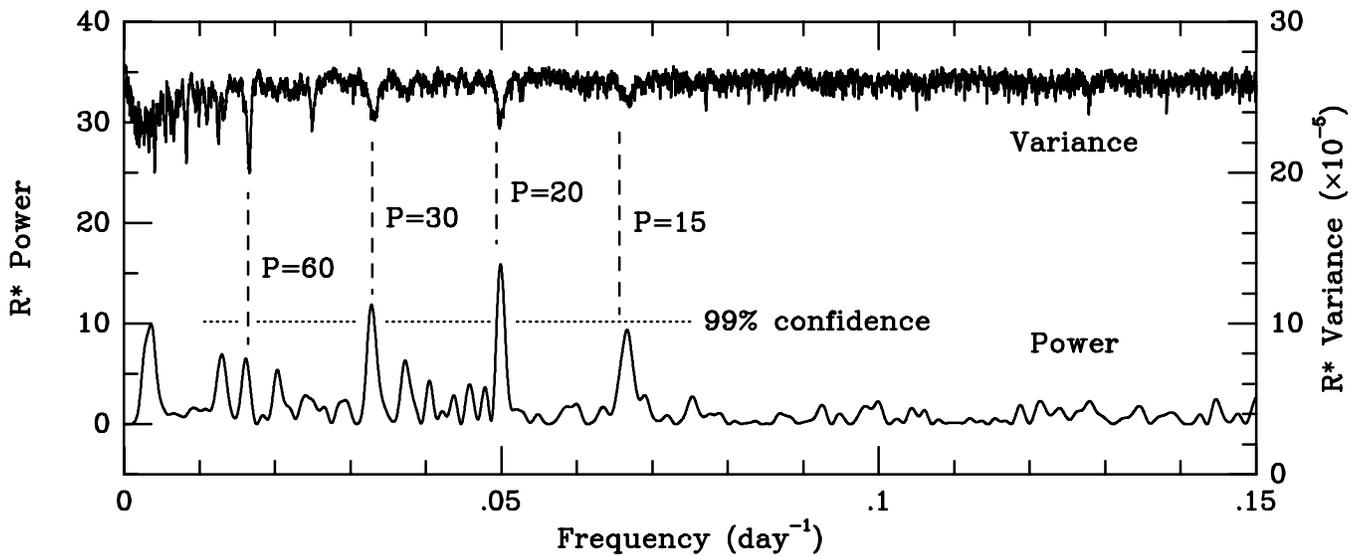
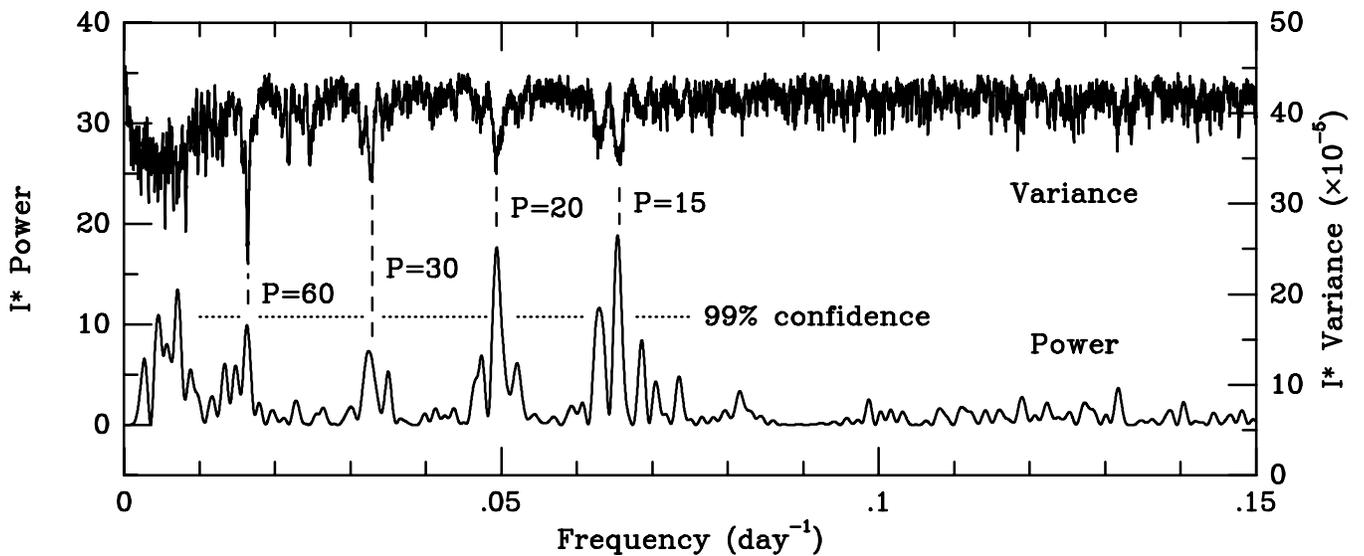
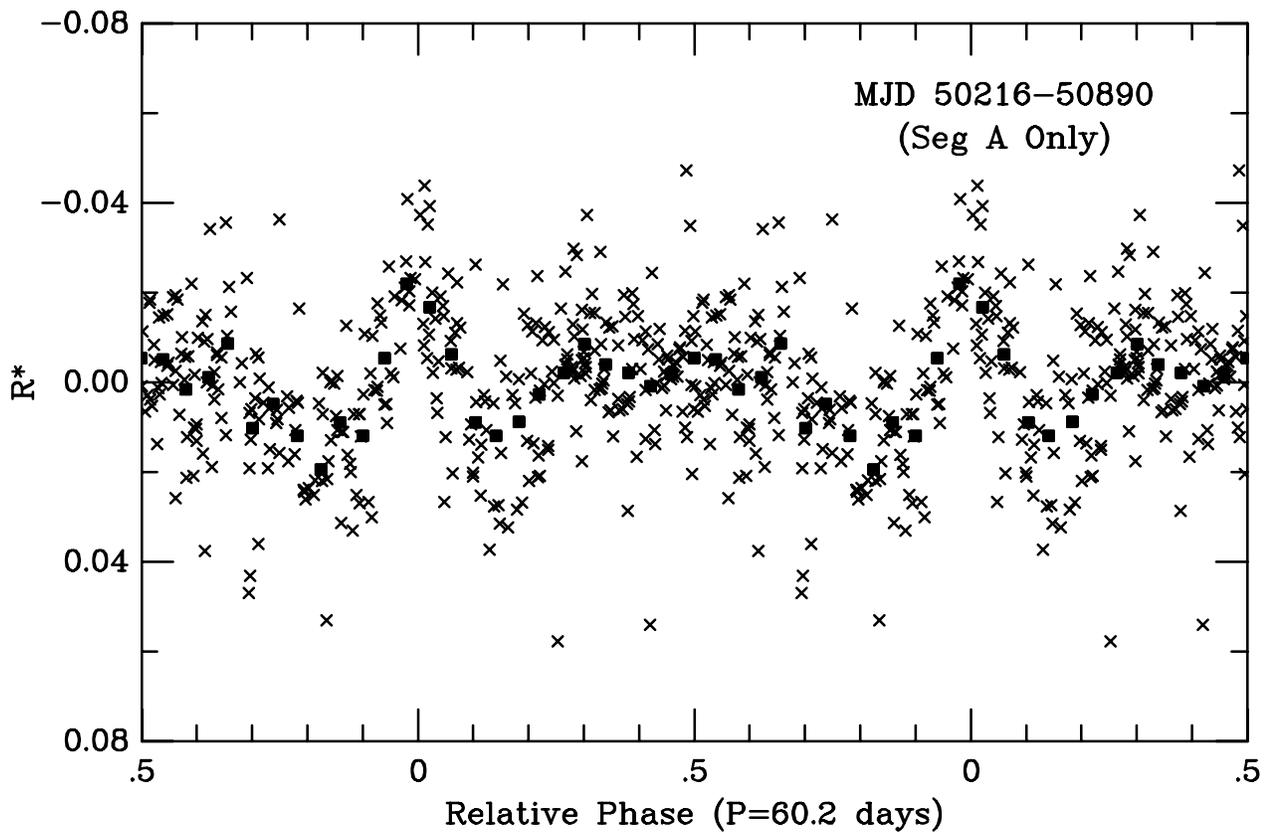
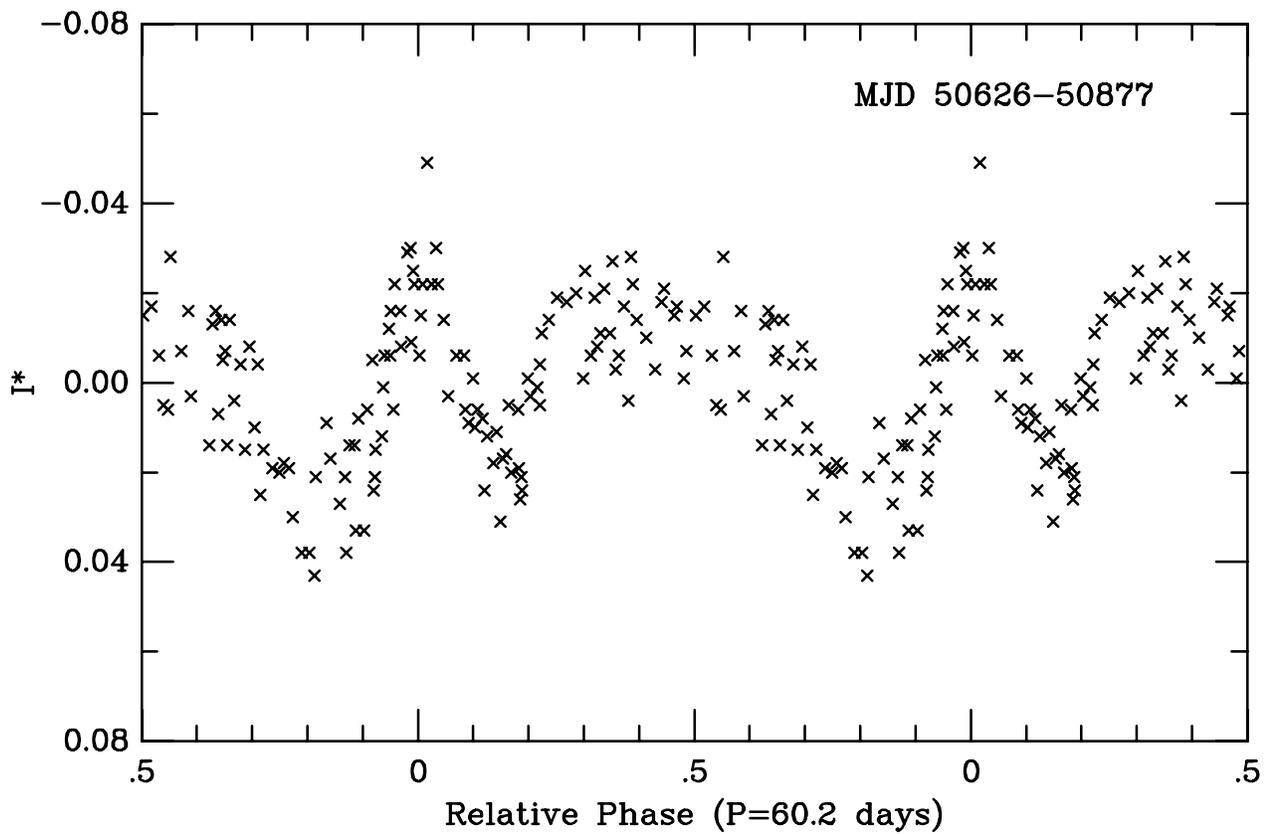
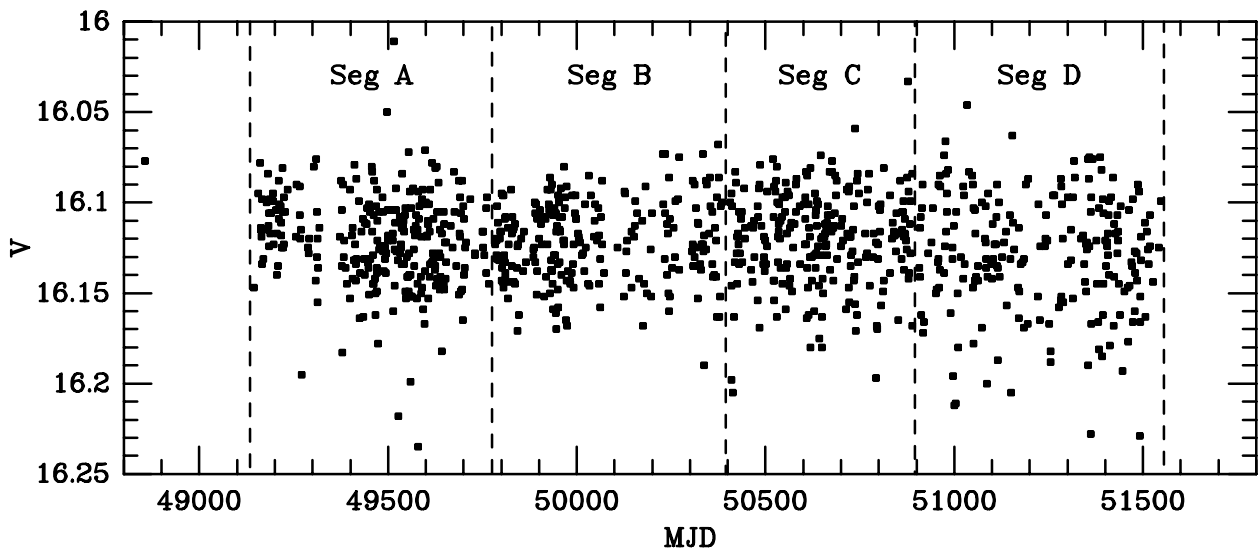
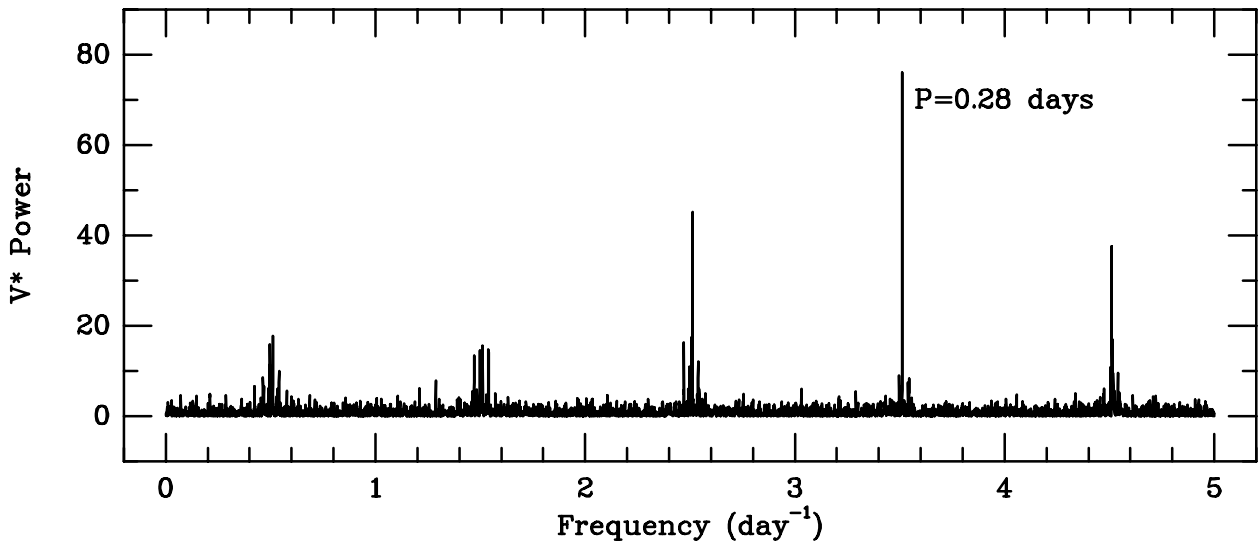
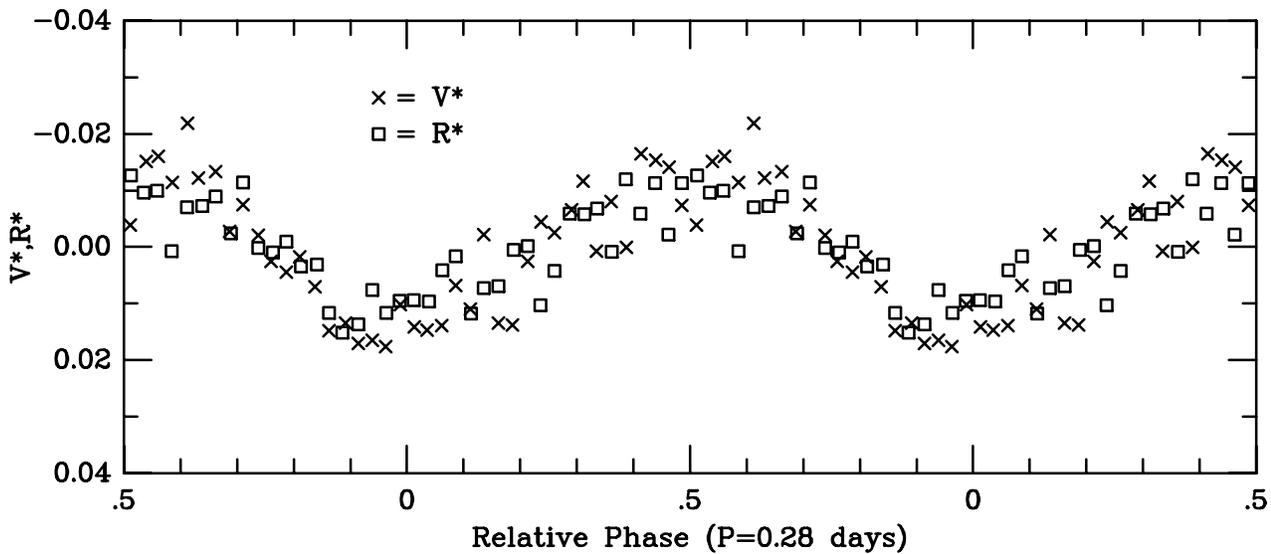
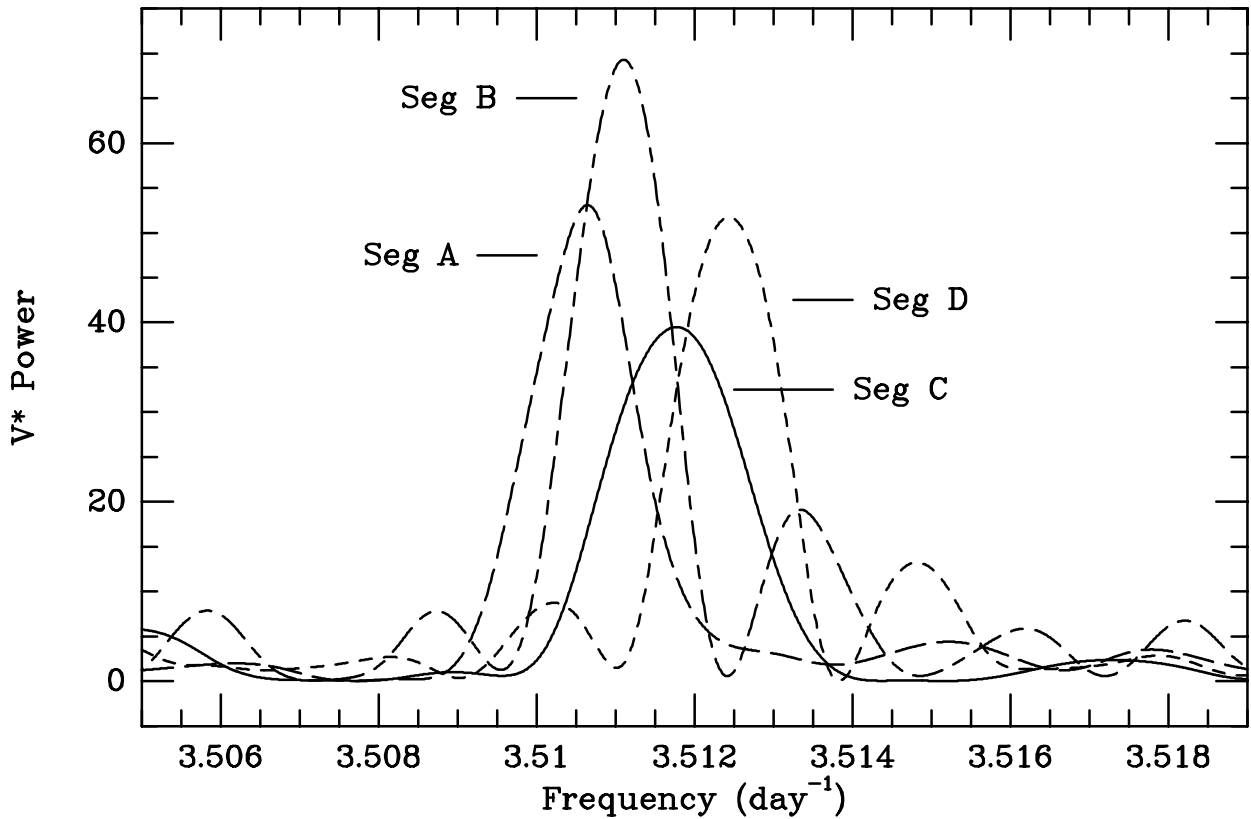
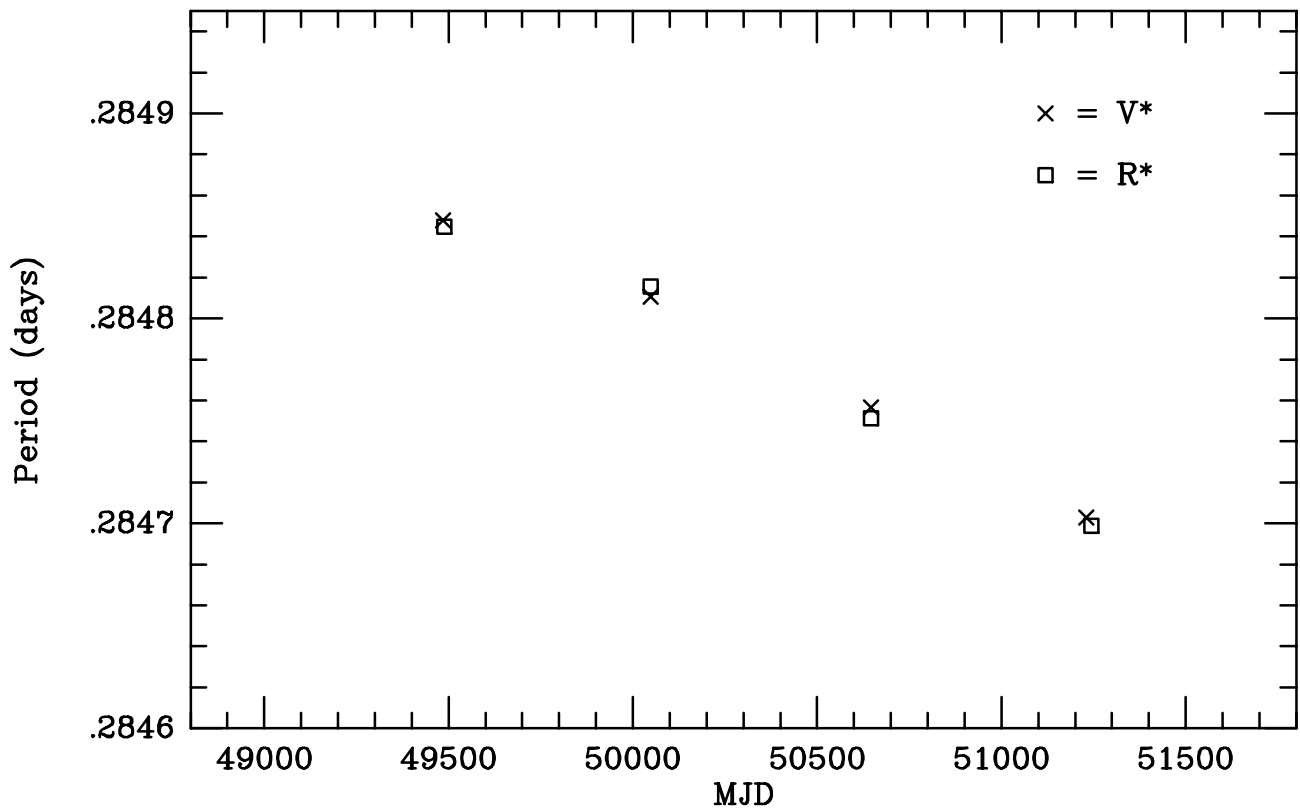
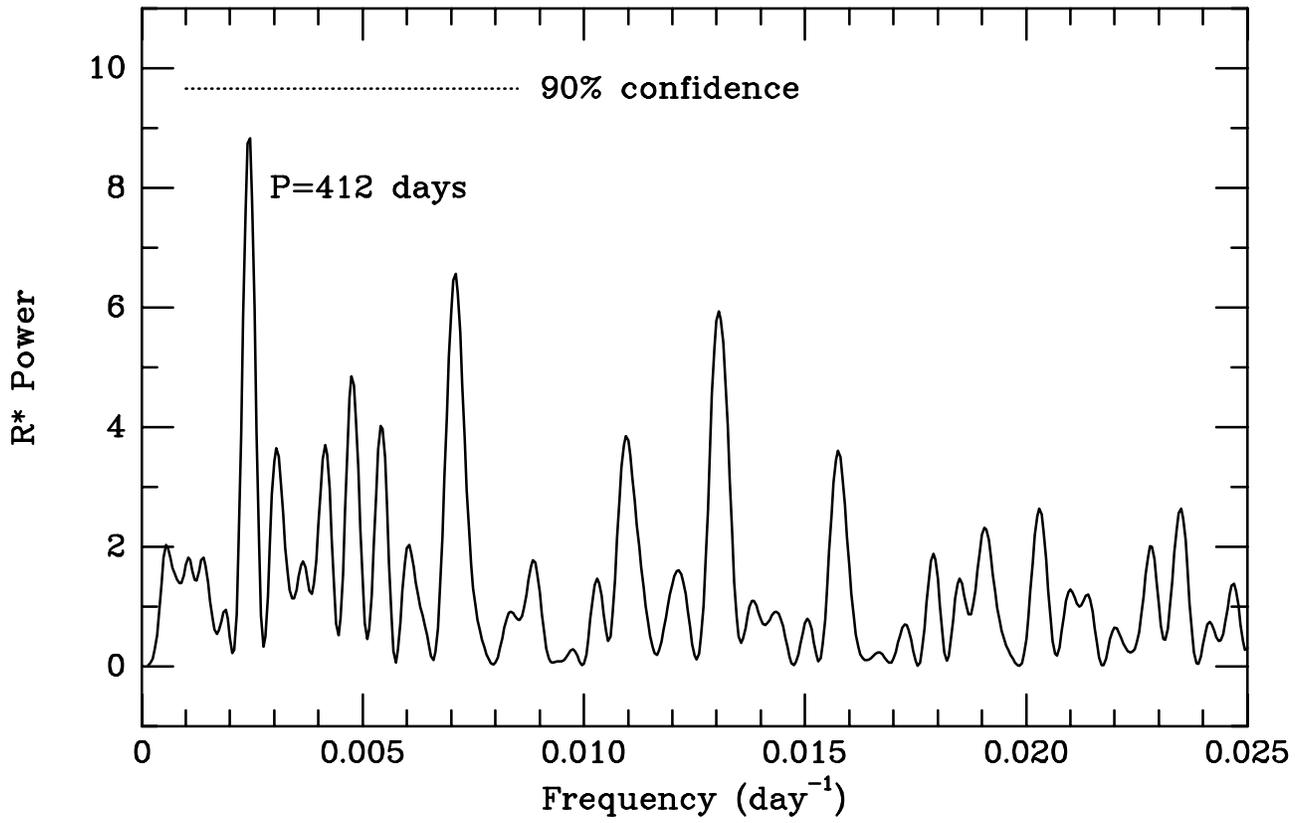
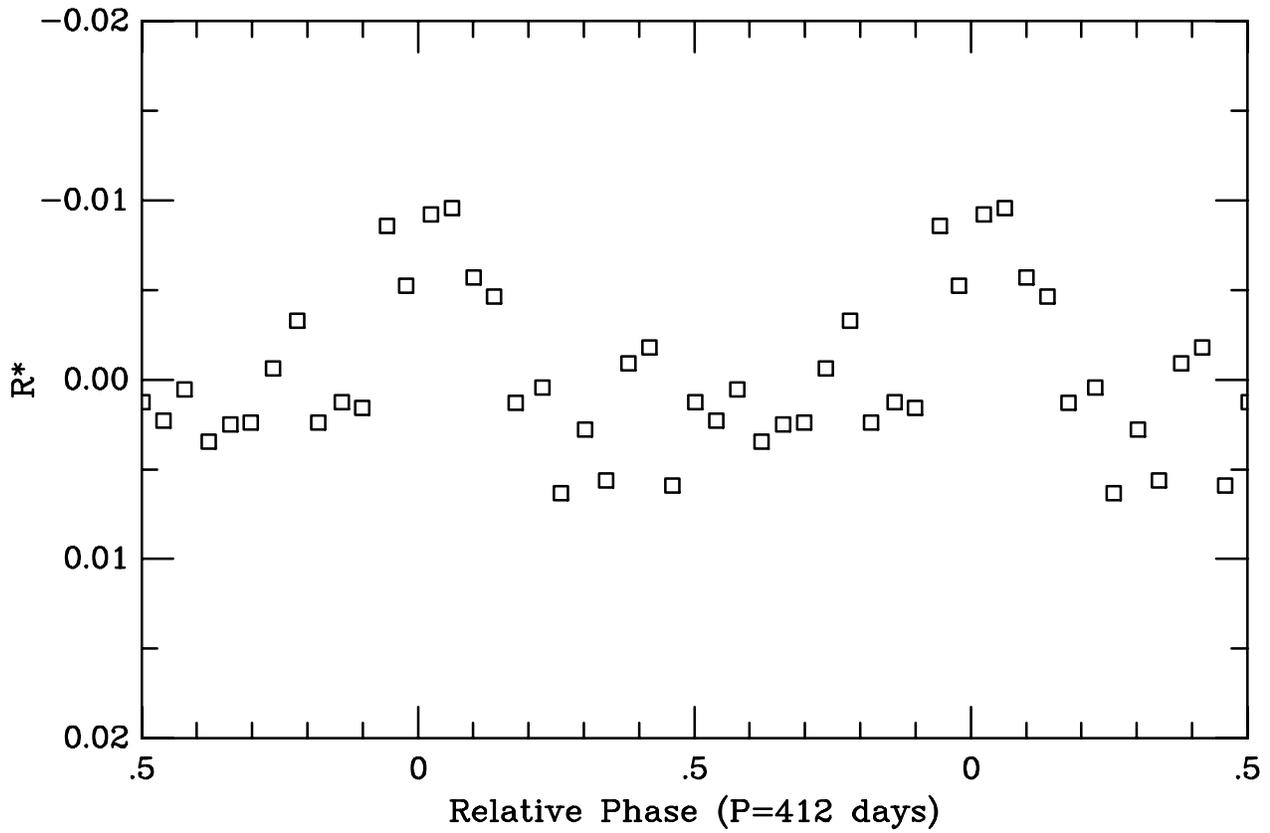
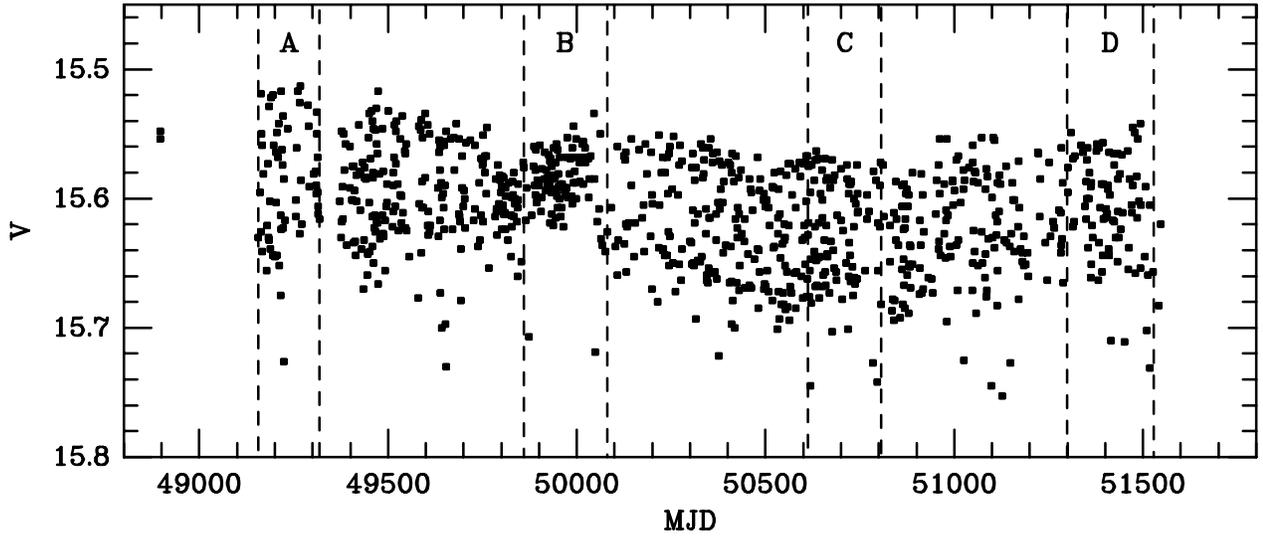
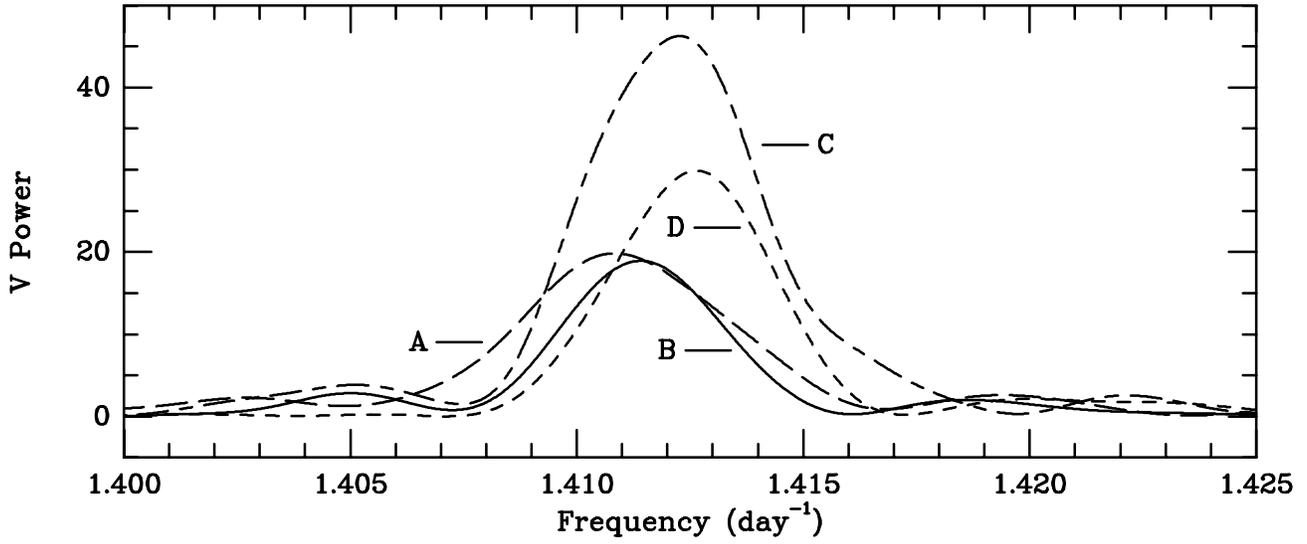
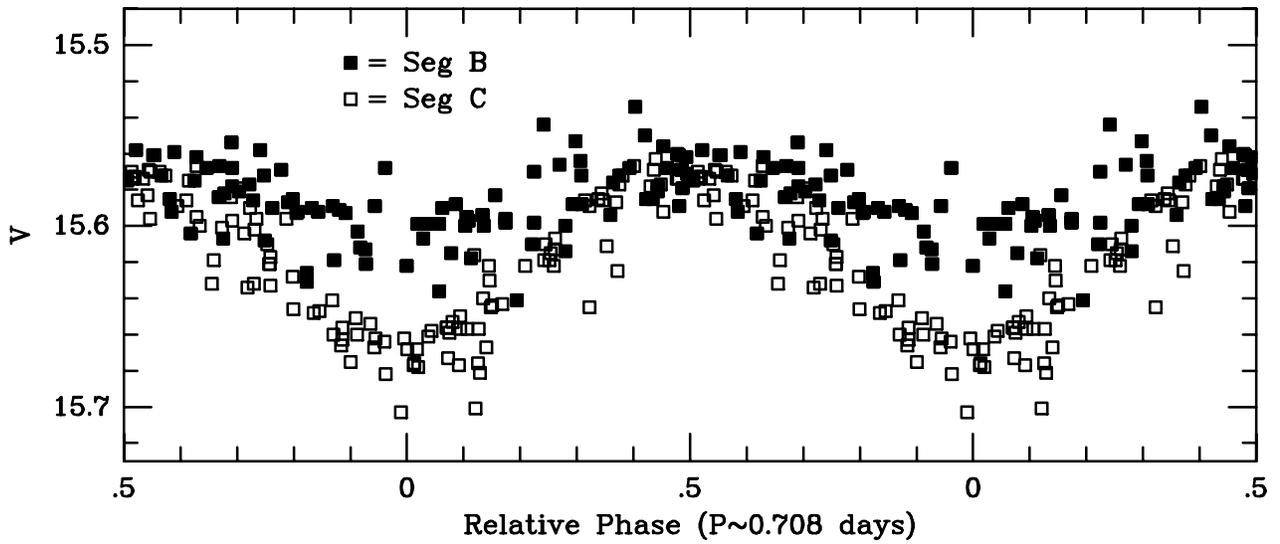
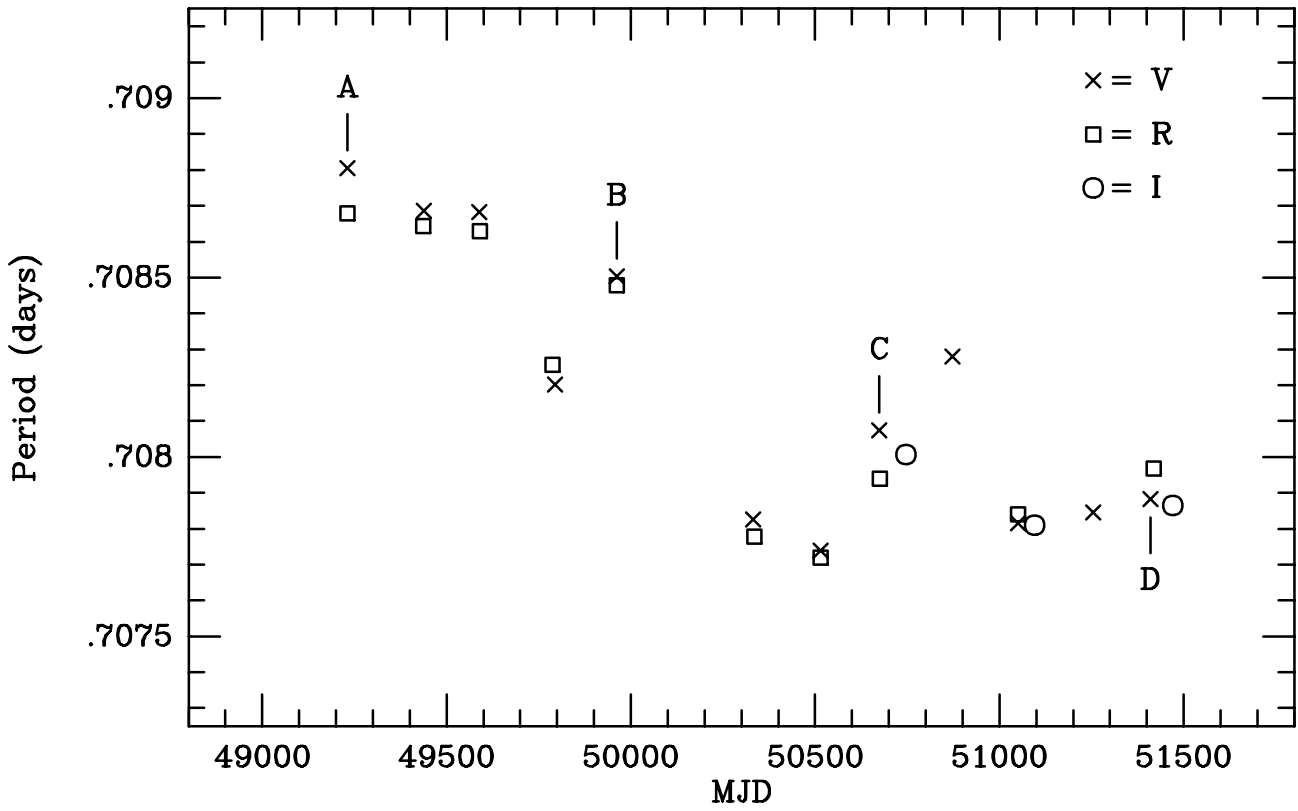
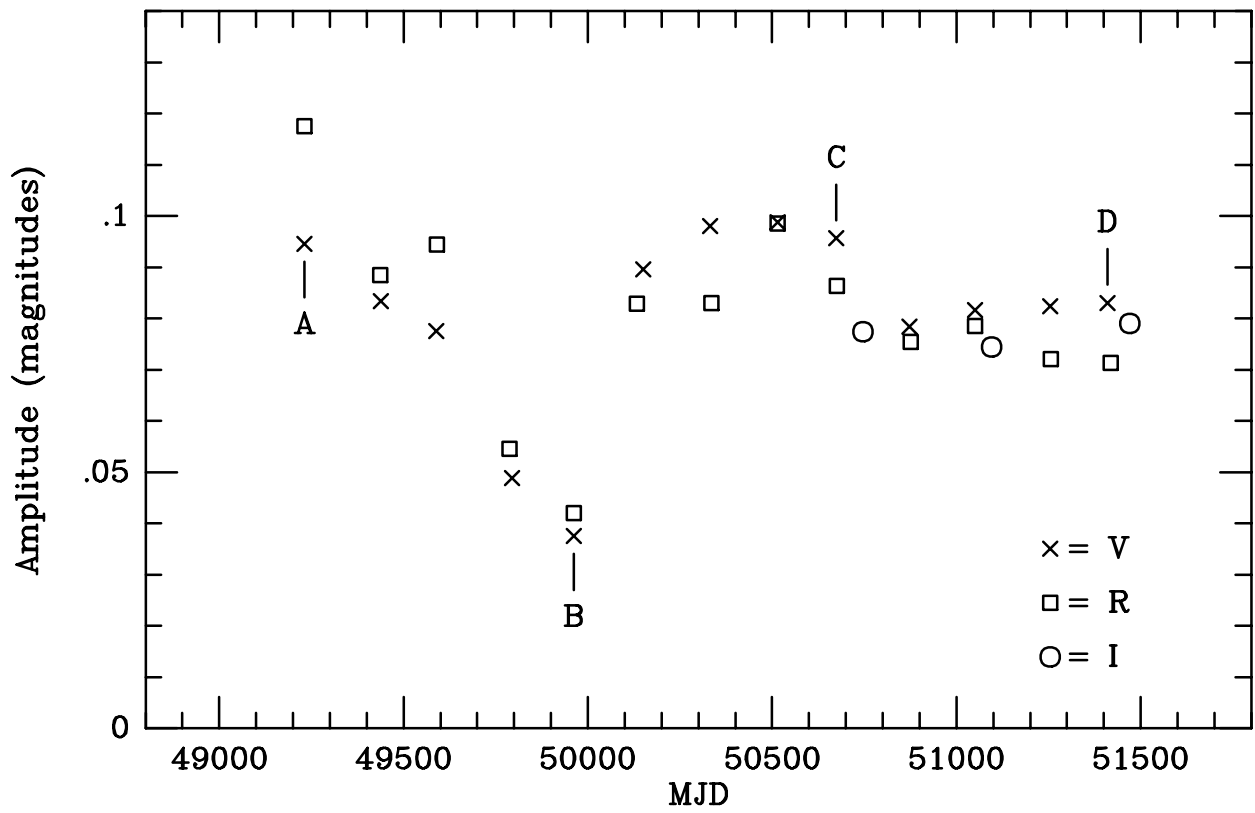
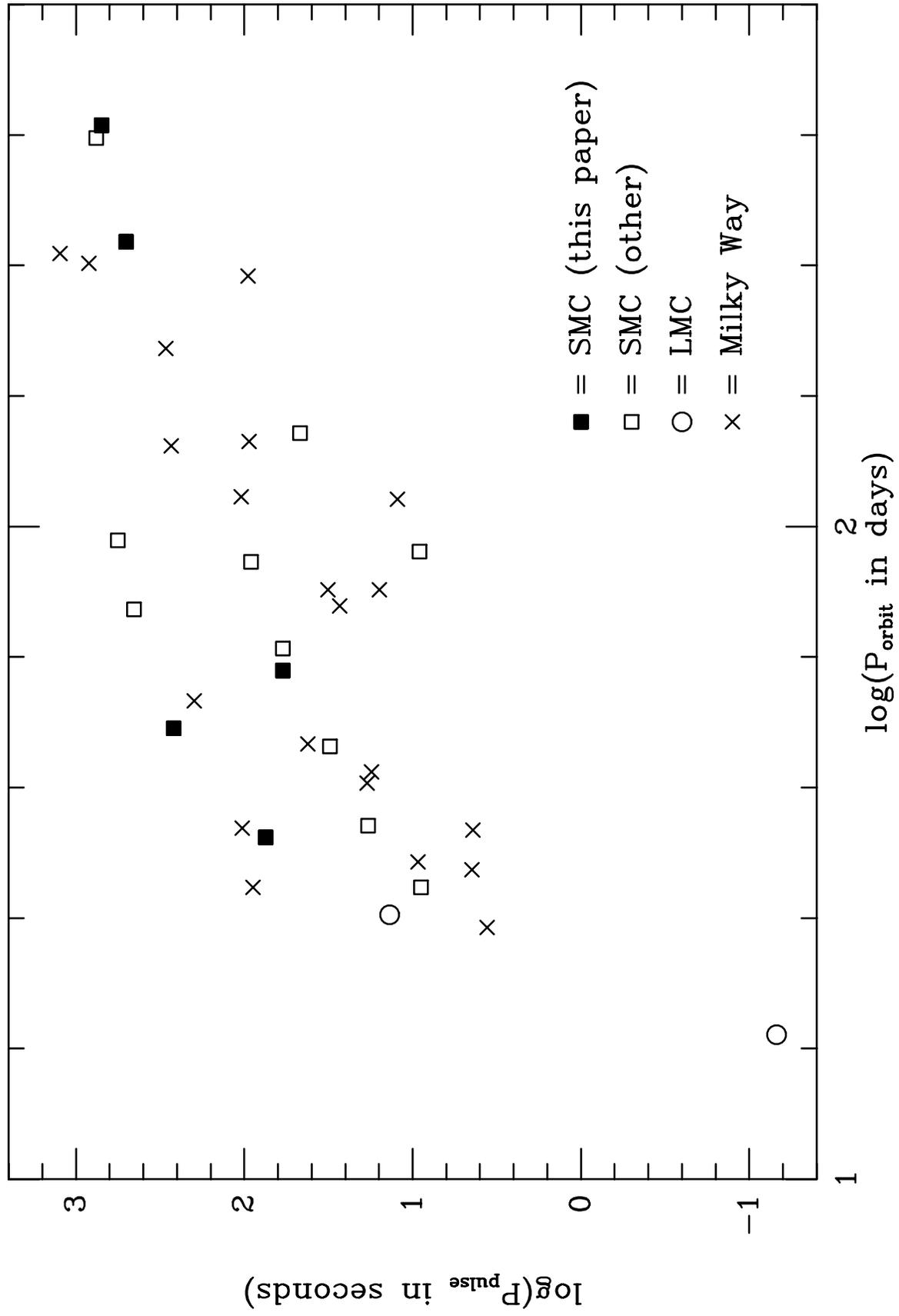
\end{document}